\documentclass[
reprint,
 amsmath,amssymb,
 aps,
]{revtex4-1}

\usepackage{graphicx}
\usepackage{dcolumn}
\usepackage{bm}

\begin{document}

\preprint{APS/123-QED}

\title{Measurement of the 10 keV resonance in the  $^{10}$B($p, \alpha_0$)$^7$Be reaction via the Trojan Horse Method}

\author{C. Spitaleri}
\email{e-mail: spitaleri@lns.infn.it}
\author{L. Lamia}
\author{S.M.R. Puglia}%
\author{S. Romano}
\author{M. La Cognata}
\author{V. Crucill\`{a}}
\author{R.G. Pizzone}
\author{G.G. Rapisarda}
\author{M.L. Sergi}

\affiliation{%
Dipartimento di Fisica e Astronomia, Universit\`a  di Catania, Catania, Italy\\
INFN, Laboratori Nazionali del Sud, Catania, Italy
 }%


\author{M. Gimenez Del Santo}
\author{N. Carlin}
\author{M. G. Munhoz}
\author{F.A.Souza}
\author{A.Szanto de Toledo}
\affiliation{%
Departamento de Fisica Nuclear, Universitade de Sao Paulo, Sao Paulo, Brasil
}%

\author{A. Tumino}

\affiliation{%
Universit\`a degli Studi di Enna ''Kore'', Enna, Italy\\
INFN, Laboratori Nazionali del Sud, Catania, Italy
 }%

\author{B. Irgaziev}
\affiliation{%
GIK Institute of Engineering Sciences and Technology, Topi, Districti Swabi, K.P. Pakistan
}%


\author{A. Mukhamedzhanov}
\author{G. Tabacaru}
\affiliation{%
 Cyclotron Institute, Texas A\&M University, College Station, TX 77843, USA
}%

\author{V. Burjan}
\author{V. Kroha}
\author{Z. Hons}
\author{J. Mrazek}
\affiliation{Nuclear Physics Institute of ASCR, Rez, Czech Republic
}%

\author{Shu-Hua Zhou}
\affiliation{China Institute of Atomic Energy, Beijing, China
}%

\author{Chengbo Li}
\affiliation{Beijing Radiation Center, 100875, Beijing, China
}%
\author{Qungang Wen}
\affiliation{Auhui University, 230601, Hefei, China
}%

\author{Y. Wakabayashi}
\author{H. Yamaguchi}
\affiliation{ Center for Nuclear Study, University of Tokyo, RIKEN Campus, 2-1 Hirosawa, Wako, Saitama, 351-0198, Japan
}

\author{E. Somorjai}
\affiliation{Atomki, Debrecen, Hungary
}


\date{\today}

\begin{abstract}
The $^{10}$B(p,$\alpha_0$)$^7$Be bare nucleus astrophysical S(E)-factor has been measured for the first time at energies from about 100 keV down to about 5 keV by means of the Trojan Horse Method (THM). In this energy region, the S(E)-factor is strongly dominated by the 8.699 MeV $^{11}$C level (J$^{\pi}$=$\frac{5}{2}$$^+$), producing an s-wave resonance centered at about 10 keV in the entrance channel. Up to now, only the high energy tail of this resonant  has been measured, while the low-energy trend is extrapolated from the available direct data. The THM has been applied to the quasi-free $^2$H($^{10}$B,$\alpha_0$$^7$Be)n reaction induced at a boron-beam energy of 24.5 MeV. An accurate analysis brings to the determination of the $^{10}$B(p,$\alpha_0$)$^7$Be S(E)-factor and of the corresponding electron screening potential $U_e$, thus giving for the first time an independent evaluation of it. 
\end{abstract}

\pacs{Valid PACS appear here}
\maketitle

\section{\label{sec:level1}Introduction}
Boron depleting reactions play an important role in understanding different  scenarios, ranging from astrophysics to applied nuclear physics. In particular, the measurements of  (p,$\alpha$) reactions on boron,  beryllium and lithium isotopes are of particular interest to determine light element abundances in stars. These elements are destroyed at different depths in  stellar interiors and  residual (atmospheric) abundances can be used  to constrain mixing phenomena occurring in such stars \cite{boesg}. Boron burning is triggered at temperatures T$\geq$5$\cdot$10$^6$ K  and takes place mainly through (p,$\alpha$) processes, with a Gamow peak \cite{Rolfs1988} centered at about 10 keV. In this context, the $^{10}$B(p,$\alpha_0$)$^7$Be reaction, for which $^7$Be nuclei is left in its ground-state, has special interest. Its cross section  at the  Gamow energy (E$_G$) is in fact dominated by the contribution of the 8.699 MeV  $^{11}$C level (J$^\pi$=$\frac{5}{2}^+)$, producing an s-wave resonance centered at about 10 keV.\\
As for   applied nuclear physics, proton-induced reactions on natural boron  $^{nat}$B, containing $^{11}$B ($\sim$80$\%$) and $^{10}$B ($\sim$ 20$\%$), have been considered as possible  candidates for  ``clean-fusion"  processes for energy production  \cite{Peterson1975}. However, since the $^{10}$B(p,$\alpha_0$)$^7$Be reaction is  origin of  radioactive fuel contamination by $^7$Be, its cross-section must be precisely known at typical energies $\leq$100 keV, where the resonant contribution strongly influences the cross-section behavior.\\
However, direct cross section measurements at ultra low energies are extremely difficult to be performed, mainly because of  the Coulomb barrier penetrability that reduces the cross section to values as small as few $picobarn$ \cite{Rolfs1988}  and because of the electron screening effects \cite{Assenbaum1987, Strieder2001}. Thus, a direct evaluation of the cross section $\sigma(E)$  is  severely hindered and beyond the present technical possibilities. To obtain the cross section value $\sigma(E_G)$  at the Gamow energy,  extrapolation should be used.  But cross sections at ultra-low energy experience variations  of many orders of magnitude  making extrapolation difficult and often unreliable.

\begin{figure}
\begin{center}
\includegraphics[scale=0.50]{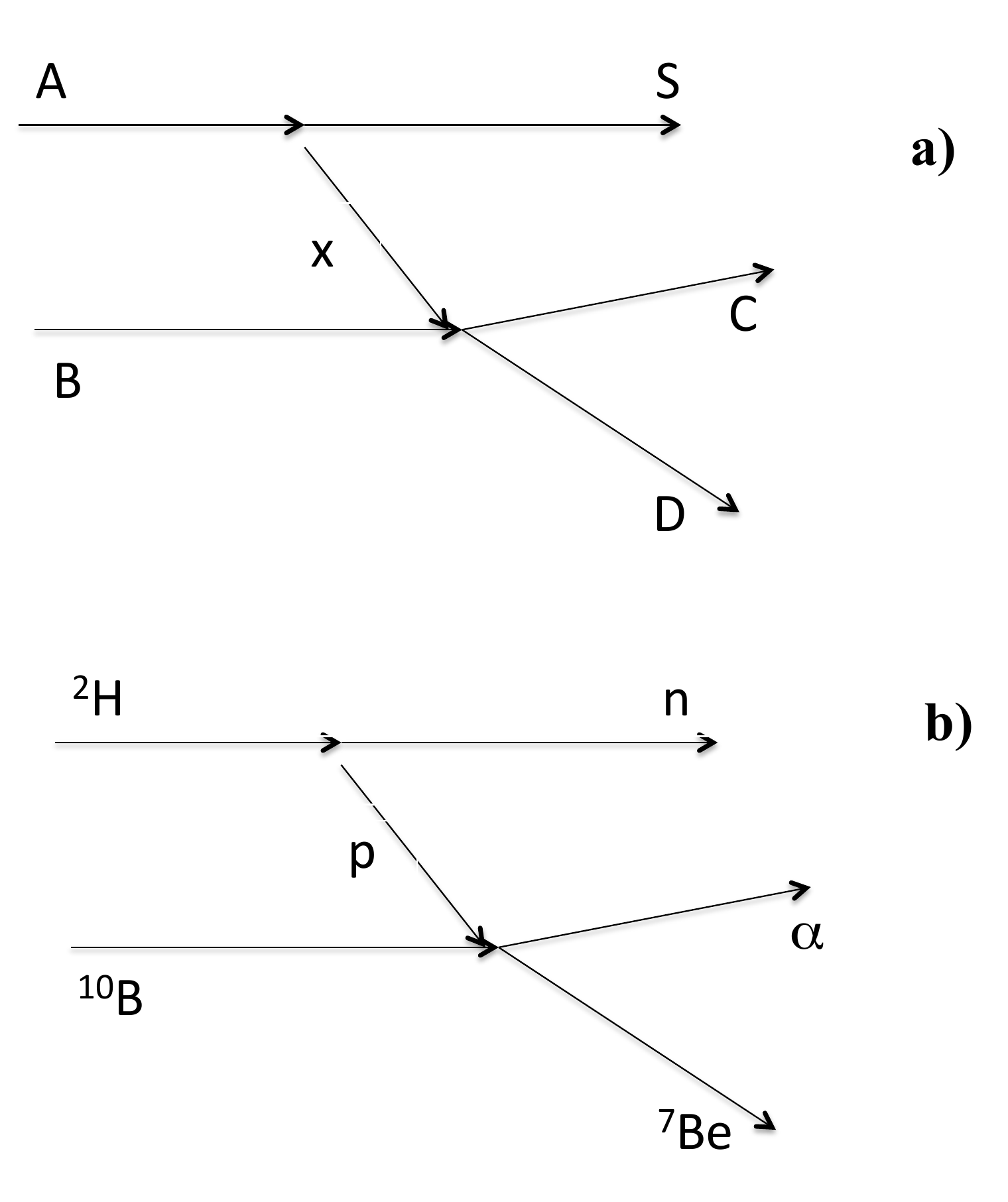}
\caption{ a) Diagram representing the quasi-free process  A + B $\to$ C + D + S. The upper vertex describes the $virtual$ $decay$ of the THM-nucleus $A$ into the clusters $x$ ($participant$) and $S$ ($spectator$); the cluster S is considered to be spectator to the x+B $\to$ C+D reaction that takes place in the lower vertex. b) Schematic diagram for the quasi-free reaction  $^2$H + $^{10}$B $\to$$ \alpha$ + $^7$Be +n.}
\end{center}
\end{figure}

To remove the  strong energy dependence due to Coulomb barrier penetration,  the  astrophysical S(E)-factor is introduced via the  relation:
\begin{equation}
S(E) = {E} \cdot \sigma(E)\cdot {exp(2\pi\eta)}
\end{equation}
where $E$ is the center of mass energy, $\eta$ is the Sommerfeld parameter 
\begin{equation}
\eta=\frac{Z_1Z_2e^2}{\hbar v}
\end{equation}
where $Z_1$ and $Z_2$ represent the charges of interacting nuclei, $v$ is their relative velocity and $exp(2\pi\eta)$ is the reciprocal of the Gamow factor.\par 
The introduction of the astrophysical S(E)-factor allows for a more accurate extrapolation procedure, especially in absence of resonances \cite{Rolfs1988}.\par
In  the $^{10}$B($p,\alpha_0$)$^7$Be case, the available direct experimental data, which are reported in the NACRE compilation \cite{Angulo1999}  and in Refs.\cite{Youn1991, Szabo1972, Wiescher1983,Angulo1993,Rauscher1996,kafkarkou2013}, refer to different experiments and range from more than 2 MeV down to about 20 keV. 
At low energies, i.e. E$<$100 keV, these data show an enhancement of the S(E)-factor due to the interplay between the 10 keV resonance and the electron screening effects \cite{Assenbaum1987, Strieder2001}. 
In addition,  no information is available  on the influence of  the tail of the sub-threshold resonance at about -35 keV, and at energies between $\sim$20 keV and $\sim$2 MeV, the different data sets disagree both in energy dependence and in the absolute value \cite{Angulo1993}.
To overcome the difficulties related to the suppression of the cross section at ultra-low energies, indirect techniques have proven to be effective.\\In particular, the Trojan Horse Method (THM, \cite{Baur1986,Cherubini1996,Spitaleri1999,Typel2003,Spitaleri2004, Akram2006, Spitaleri2011,Spitaleri2011b} and references therein) provides, at present, one of the most powerful technique for measuring the energy dependence of the bare nucleus cross section down to the astrophysically relevant  energies.
The THM  allows one to extract the low-energy S(E)-factor without Coulomb suppression and electron screening effects, which strongly influence direct measurements at astrophysical energies (see \cite{Spitaleri2011,Spitaleri2011b} and references therein). \par
The present paper reports on the first measurement of the $^{10}$B(p,$ \alpha_0$)$^7$Be S(E)-factor at $\sim$10 keV via THM, i.e. in the Gamow window for typical boron burning stellar environments. 

\section{ THE TROJAN HORSE METHOD: BASIC  THEORY}

The THM has been successfully applied to measure  the bare nucleus cross sections of  several reactions related to fundamental astrophysical and nuclear physics problems \cite{Baur1986,Cherubini1996,Spitaleri1999,Typel2003,Spitaleri2004, Akram2006, Spitaleri2011,Spitaleri2011b,Sergi2010,Gulino2012,Lacognata2011,Lamia2012,Lamia2012aa,Lamia2013apj,Lamia2007,Lamia2008, DelSanto2008,Lacognata2005,Lacognata2010,Spitaleri01,Tumino2003,Tumino06,Tumino2011,Lacognata07,Tumino2007,Tumino2008,Mukhamedzhanov2008,Lacognata2009}. Here we shortly summarize the main features of the method.

\subsection{Quasi-free reaction mechanism}
The quasi-free  (QF) A+ B $\to$ C + D + S reaction  can be described by means of the Feynman diagram shown in Fig.1 a), where only the  first term of the Feynman series is retained.
This can be described  as a transfer to the continuum, in which the nucleus A (so called TH-nucleus) breaks-up into the transferred cluster $x$ (participant) and the cluster $S$ acting  as a spectator  to the  x + B $\to$ C+ D virtual reaction. The nucleus $A$ should have a strong $x+S$ cluster structure to maximize the QF breakup yield.\\
When this reaction mechanism is present,  it can be distinguishable from others  in a region of the three body phase space where the inter-cluster momentum  ($p_{x-S}$) of the spectator $S$ is small i.e. for QF conditions.\\
The THM has its background in the theory of direct reactions (see e.g. \cite{Satchler1983}), and in particular in the studies of the QF reaction mechanisms \cite{Spitaleri1990}. The application to nuclear  reactions of astrophysical interest  is an extension to  low energies of the well-assessed measurements of QF reactions at higher energies \cite{Zadro1989,Calvi1990,Spitaleri1990}.\\
In the present application,  the QF contribution to the three-body  $^2$H($^{10}$B,$\alpha_0$$^7$Be)n reaction of Fig.1 b) \cite{Baur1986,Spitaleri1990}, performed at energy well above the Coulomb barrier in the entrance $^2$H+$^{10}$B channel, is selected to extract the $^{10}$B(p,$\alpha_0$)$^7$Be  cross section at astrophysical energies.\\
The THM is applied here within  the Plane Wave Impulse Approximation (PWIA) framework, and the motivations for such a simplified approach in the application of the THM have been discussed in \cite{Spitaleri2011,Spitaleri2011b}. Some of the  critical points of this simplified approximation are presented.\\
The QF $^2$H($^{10}$B,$\alpha_0$ $^7$Be)n reaction  can be described by the Feynman diagram (Fig.1b) \cite{Shapiro1965, Shapiro1967a,Shapiro1967}. This diagram represents the dominant process (pole approximation), while other graphs (triangle graphs) indicating re-scattering between the reaction products, are neglected \cite{Shapiro1967}. Under these hypotheses, the incident particle $^{10}$B is considered to interact only with the proton in the target nucleus $^2$H, while the neutron is considered spectator of the $^{10}$B(p,$\alpha_0$)$^7$Be virtual reaction of interest for astrophysics.\\
Following the simple PWIA, the three-body reaction cross section can be factorized into two terms corresponding to the  two vertices of Fig.1 b) and it is given by \cite{Spitaleri2011,Spitaleri2011b}:
\begin{equation}
\frac{d^3\sigma}{d\Omega_\alpha d\Omega_{^7Be} dE_\alpha}
 \propto KF \cdot \mid\Phi(\vec{p_{n}})\mid^2 \cdot 
 \left(\frac{d\sigma}{d\Omega}\right) 
 ^{HOES}
\end{equation}			      		    
where:
\begin{itemize}
\item $KF$ is a kinematical factor containing the  final state phase-space factor and it is a function of the masses, momenta and emission angles  of the two detected particles ${\alpha}$ and $^7$Be, of the incident $^{10}$B particle momentum, and of the mass of the spectator $n$. Referring to Fig.1 a), its final expression is:
\begin{eqnarray}
 KF = \frac{\mu_{AB}m_{D}}{(2\pi)^{5}\hbar^{7}}
 \frac{p_{C}p_{D}^{3}}{p_{AB}}
 \left[\left( \frac{\vec{p}_{Ys}}{\mu_{Ys}}
 - \frac{\vec{p}_{CD}}{m_{D}}\right) \cdot
 \frac{\vec{p}_{D}}{p_{D}}  \right]^{-1}
\end{eqnarray} 
where $Y$ stands for the $C+D$ system \cite{Spitaleri01};
\item $\left(\frac{d\sigma}{d\Omega}\right)^{HOES}$  
 is the half-off-energy-shell  (HOES) differential cross section for the  $^{10}$B(p,$\alpha_0$)$^7$Be 
 reaction at the center-of-mass energy  $E$,  given in post-collision prescription
  by the relation \cite{Grossiodor1974}
\begin{equation}
E = E_{\alpha-^7Be}-Q                     
\end{equation}
where $Q$ is the  $Q$-value  for the $^{10}$B(p,$\alpha_0$)$^7$Be  reaction  and E$_{\alpha-^7Be}$  is the  ${\alpha-^7Be}$ relative energy.
\item ${\Phi}(\vec{p_n})$ is the Fourier transform of the radial wave function $\chi(\vec{r}_{pn})$ of the $p-n$ inter-cluster motion, usually given by the Hulth\'en function. 
\end{itemize}
In the deuteron, the $p-n$ relative motion is most likely taking place  in $s$ wave, thus the  momentum distribution has a maximum at $p_n$= 0 MeV/c  ($p_n$ is the inter-cluster momentum $\equiv$  $p_{x-S}$).\\More sophisticated theoretical formulations, accounting for HOES effects and the spin-parity of the interacting nuclei, can be found in \cite{Typel2000,Typel2003, Akram2006,Mukhamedzhanov2008,Akram2011}.

\subsection{Energy and momentum prescriptions}
The beam energy has to be carefully chosen to span the Gamow window under QF conditions.\par
Moreover,  the validity conditions of the Impulse Approximation  ($IA$) were checked. Since the $^{10}$B incident energy of 24.5 MeV  corresponds to a quite high momentum transfer $q_t$= 220 MeV/c \cite{Barbarino1980,Pizzone2005,Pizzone2009} and  to an associated de Broglie wavelength  $\lambda$= 0.89  fm, smaller enough with respect the deuteron effective radius of about 4.5 fm \cite{Chew1952}, it is expected that the Impulse Approximation (IA) represents a suitable description of the process. This will be verified during the data analysis.\par
 The beam energy  (equal to about 4.1 MeV in the center-of-mass system) is large enough to overcome the Coulomb barrier $V_{C}$ = 1.62 MeV in the entry $^2$H+$^{10}$B channel. Thus, the proton is brought inside the nuclear field of $^{10}$B to induce the $^{10}$B+ p $\to$ $\alpha_0$+ $^7$Be reaction. \par
Even if the beam energy was much larger than the in direct experiments, the THM has allowed us to investigate this range.
This is possible because the initial projectile energy is compensated for by the binding energy of deuteron (\cite{Spitaleri1999, Spitaleri2004} and references therein), making the relative energy $E_{cm}$ very low. In symbols:
\begin{equation}
E_{cm} = E_{p-^{10}B} - B_{np}  
\end{equation}
where E$_{p-^{10}B}$  is the the projectile energy in the two body proton-$^{10}$B center-of-mass system and $B_{np}$ the $p-n$ binding energy. \par
 
The applicability of the $IA$  is limited to small   $p_{n}$ momenta, satisfying the condition given in Ref. \cite{Shapiro1967a}
\begin{equation}
 p_n \leq k_n
\end{equation}
where 
$k_n$ = $\sqrt{2 \mu_{np} B_{np}}$ and $\mu_{np}$   the p-n reduced mass. 
For deuterons,  the limit  (6) is;
\begin{equation}
 p_n \leq 44 MeV/c 
\end{equation}

\section{THE EXPERIMENT}

\subsection{Selection of the Trojan horse nucleus}
The $^{10}$B(p,$\alpha_0$)$^7$Be cross section measurement can be performed using of a participant proton hidden either inside a deuteron $^2$H =$(p + n)$ with $n$ = spectator (binding energy $B_{pn}$= 2.225 MeV) or inside  $^3{\rm He}$ =$(p + d)$  with $d$ = spectator ($B_{pd}$= 5.49 MeV).  
The spectator-particle independence of the cross section has been proved in a number of works  \cite{Tumino2005, Pizzone2011, Pizzone2013}.
The $^2$H($^{10}$B,$\alpha_0$$^7$Be)n cross-section measurement is performed in inverse kinematics  by using deuteron target as a virtual-proton target, as already done in a large number of indirect investigations with the THM \cite{Spitaleri2011, Tumino2003, Lacognata2007, Lacognata2011, Lacognata2008, Sergi2010, Lamia2008, Lamia2012}.\par

 The choice of a deuteron as TH-nucleus is suggested by a number of reasons:
\begin{enumerate}
\item  its relatively low binding energy;
\item  its well known radial wave function;
\item  its obvious proton-neutron structure;
\item  it provides neutral spectator, if proton is chosen as participant;
\item  the $p-n$ relative motion takes place in $l=0$, thus the momentum distribution $\mid$${\Phi}(\vec{p_n})$$\mid$$^2$  has a maximum for $p_n$ = 0 MeV/c;
\item the small effects of the $d$-wave component (less than 1\% \cite{Lamia2012prc}).
\end{enumerate}

\subsection{Experimental setup}

The experiment was performed at the  Laboratori Nazionali del Sud (LNS) in Catania (Italy).  The SMP Tandem Van de Graaf accelerator provided a 24.5  MeV $^{10}$B beam with an intensity of $\sim$1.5 nA. The beam spot was reduced to 2 mm in diameter  using a collimator. An anti-scattering system was used to preserve detectors at small angles from scattered beam. The relative beam energy spread was about 10$^{-4}$. Self-supported 200 $\mu$g cm$^{-2}$  thick CD$_2$ target were placed at 90$^\circ$ with respect to the beam direction.\\
The detection setup  consisted of a $\Delta$E-E system, made up of an ionization chamber (I.C.) (as $\Delta$E stage), with mylar entrance (0.9 micron thick), and exit (1.5 micron thick) windows,  filled with butane gas at a pressure of about 40 mbar. A silicon position sensitive detector (single area, resistive redout) PSD$_A$ was used to detect the residual energy of the emitted particles.  
\begin{figure}
\begin{center}
\includegraphics[scale=0.30]{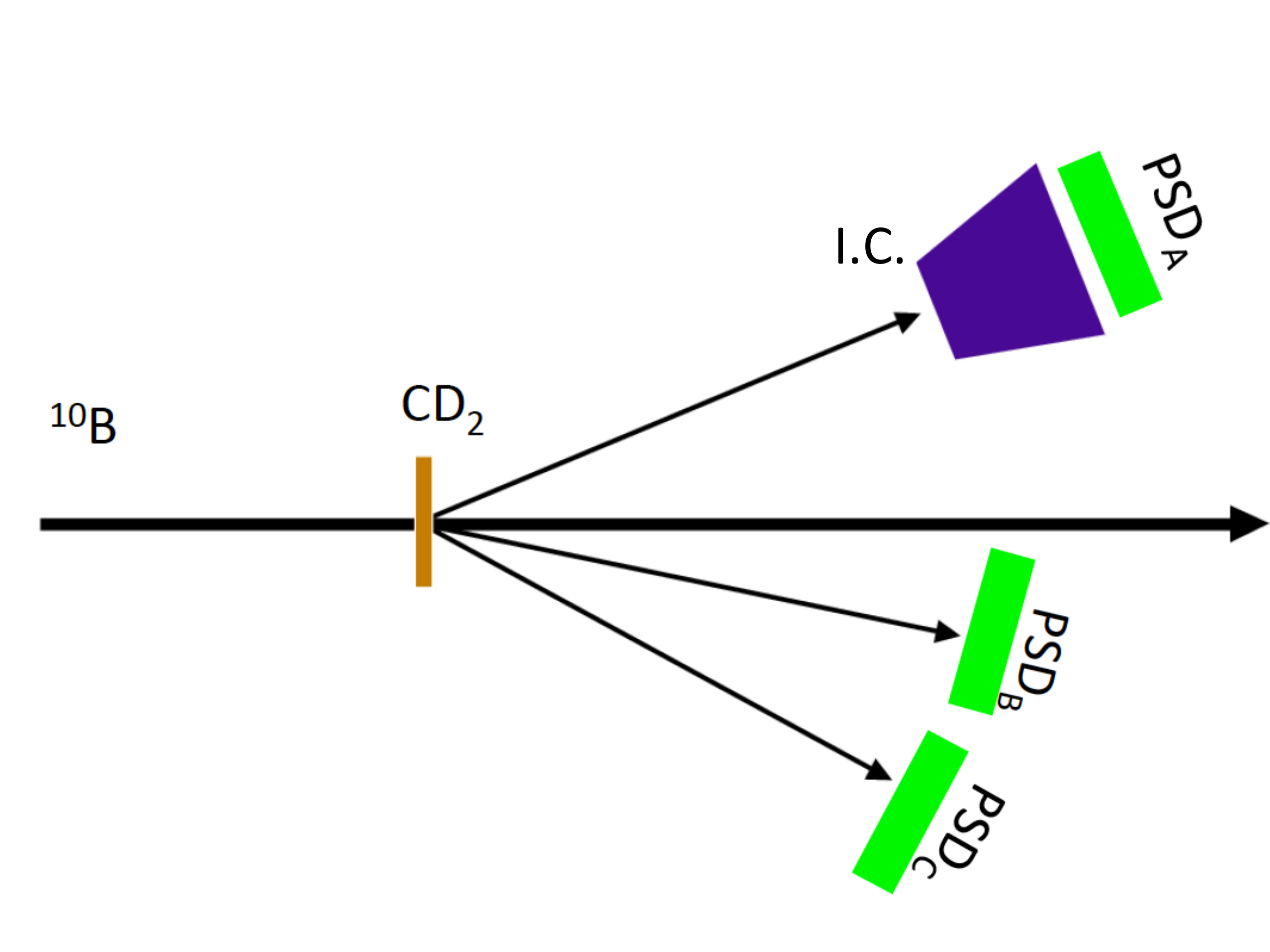}
\caption{(Color online) Schematic drawing of the adopted experimental setup, showing the $\Delta$E-E system, made up of an ionization chamber (I.C.) and a position sensitive detector (PSD$_A$), devoted to $^7$Be detection, and PSD$_B$ and PSD$_C$, devoted to alpha particle detection.}
\end{center}
\end{figure}
Two position sensitive detectors PSD$_B$ and PSD$_C$  were placed at opposite side with respect to the beam direction (Fig.2). Thanks to the diameter of the scattering chamber ($\sim$ 2000 mm), the detectors were fixed at a distance of $\sim$ 600 mm from the target. Details of the adopted experimental setup (i.e. angular position, distances, solid angles etc.) are listed in Table~\ref{setup}, together with the intrinsic angular resolution $\delta\theta$.  The coplanarity of the three detectors was checked by an optical system.\\
Angular ranges  were chosen to cover neutron momenta p$_n$ ranging from -200 MeV/c to  200 MeV/c.  This assures that the bulk of the quasi-free contribution for the breakup process of interest lies inside the investigated region. This allowed also to cross check the method inside and outside the phase-space regions where the quasi-free contribution is expected.\\
The energy and position signals of the PSDs were processed by standard electronics together with the time signals coming from any two of them. 
The trigger for the data acquisition was given by the logic coincidence between the $\Delta$E-E system and the ``OR" of logic signals from PSD$_B$ and PSD$_C$. The processed signals were then sent to the acquisition system for on-line monitoring and data storage. 
Deterioration of CD$_2$ targets has been continuously overseen by monitoring the ratio of the Z=4 particle yield to the charge collected in the Faraday cup at the end of the beam line.
\begin{figure}
\begin{center}
\includegraphics[scale=0.65]{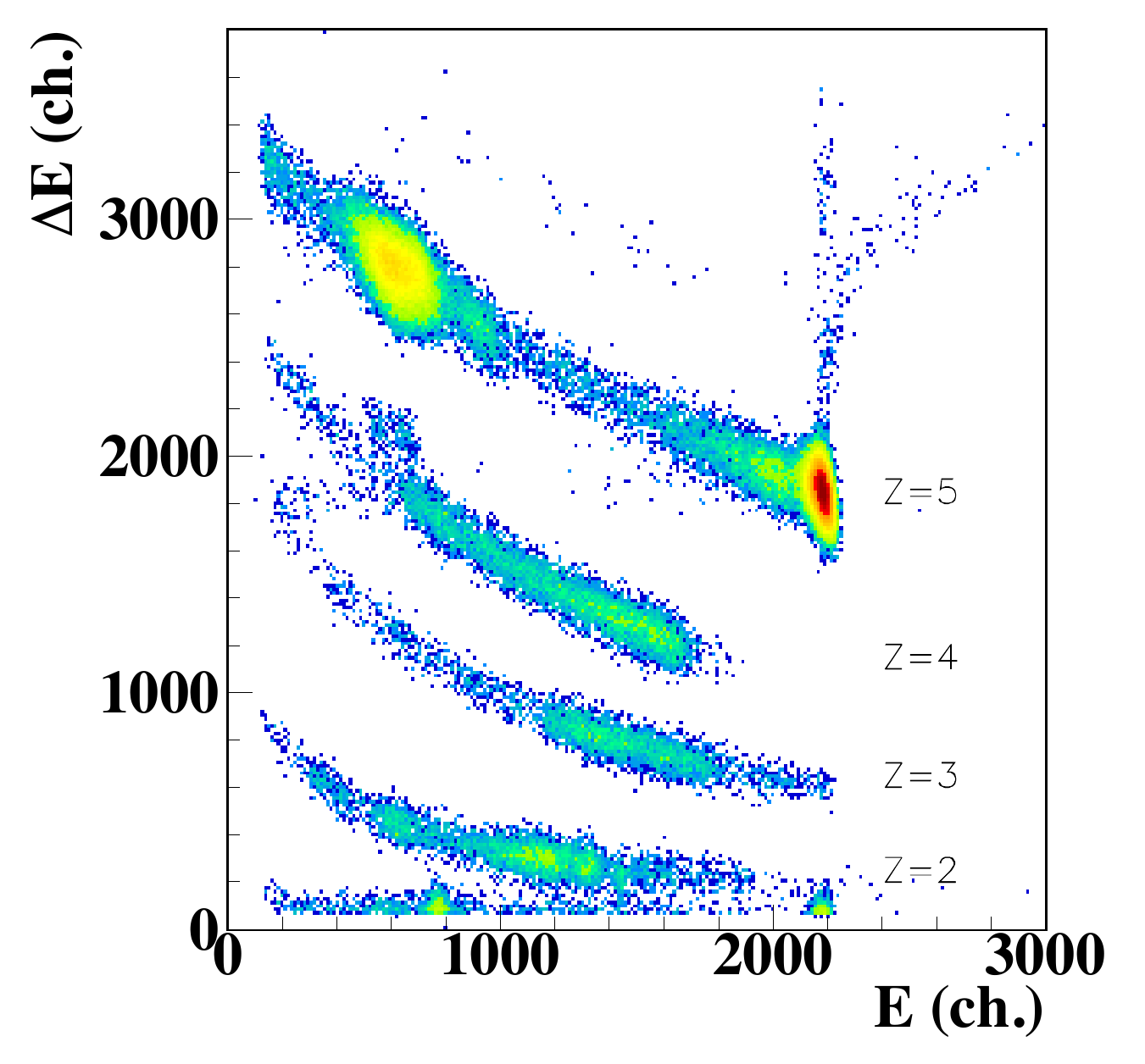}
\caption{(Color online) The 2D $\Delta$E-E plot, showing the energy loss in the ionization chamber ($\Delta$E) as a function of the residual energy detected in PSD$_A$.}
\end{center}
\end{figure}

\begin{table}[h]
\caption{Laboratory central angles ($\theta_0$), covered angular ranges ($\Delta\theta$),  solid angles ($\Delta\Omega$), distances from the target ($d$), thickness ($s$),  effective area, and intrinsic angular resolution ($\delta\theta$) for each detector.}
\begin{tabular}{lcccccccc}
\hline 
\hline 
{}&Detector  {}&$\theta_0$  {}&$\Delta \theta$  {}&$\Delta \Omega$ {}&$d$      {}&$s$                {}&Area     {}&$\delta\theta$               \\ [1.0ex]           
{}&                {}&(deg)          {}&(deg)                 {}&( msr)                  {}&(mm) {}&($\mu$m)  {}&(cm$^2$)   {}&(deg)                       \\ [1.5ex]
\hline

{}&$PSD_A$            {}&6.9        {} &5                    {}&1.5$\pm$0.1          {}&570$\pm$2     {}&492    {}&5    {}&0.10                           \\[1ex]
  
 {}&$PSD_B$             {}&8.2                     {} &8                      {}&4.1$\pm0.4$            {}&350$\pm$2         {}&492                      {}& 5           {}&0.16                       \\[1ex]
 
 {}&$PSD_C$            {}&17.9                    {} &8.6                  {}&$4.6\pm0.3$             {}&330$\pm$2         {}&984                      {}&5            {}&0.17                          \\[1ex]
\hline
\hline
\end{tabular}
\label{setup}
\end{table}
\begin{figure}
\begin{center}
\includegraphics[scale=0.60]{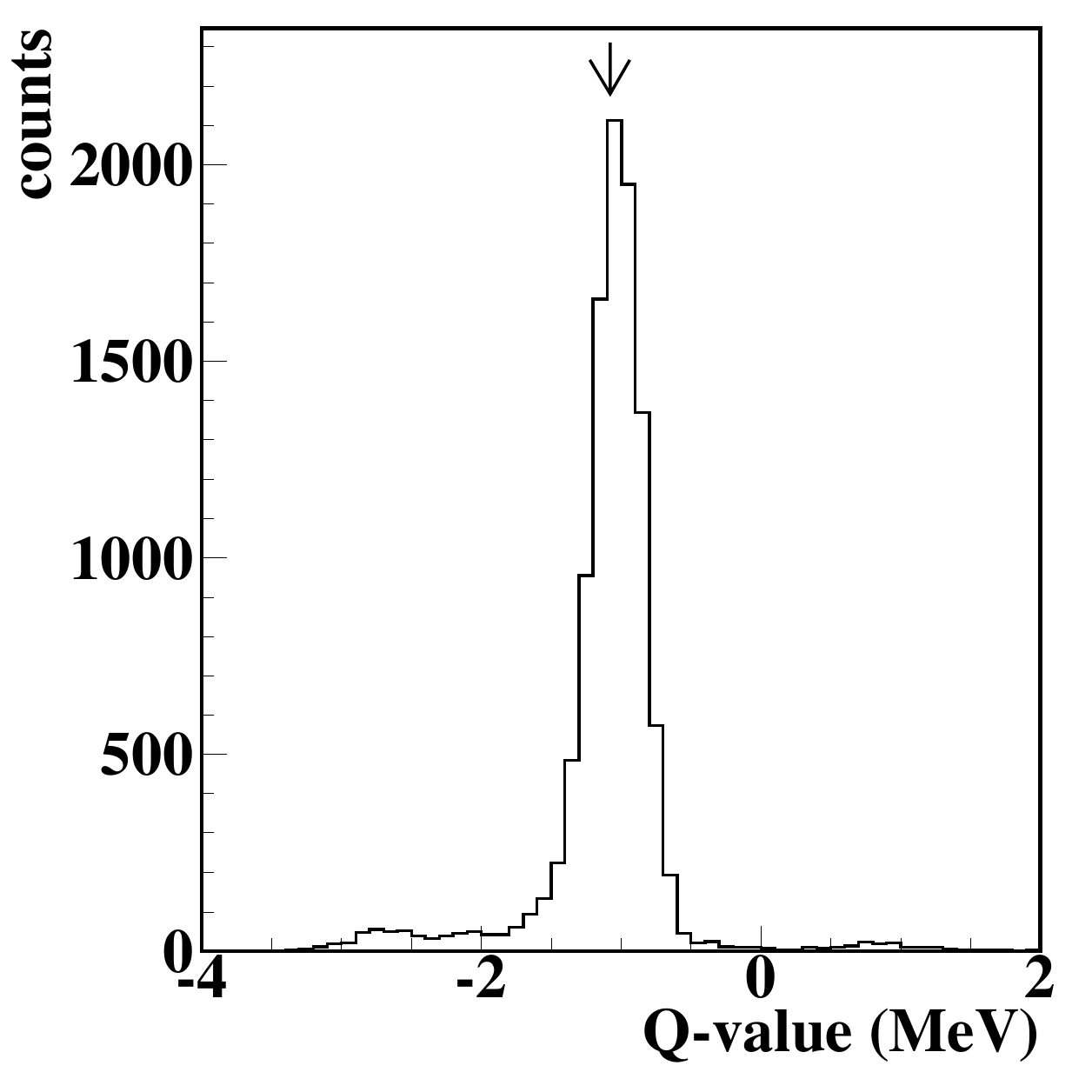}
\caption{Experimental $Q$-value spectrum.The vertical arrow marks the position of the theoretical Q value of the $^2$H($^{10}$B,$\alpha$$^7$Be)n reaction. 
No reactions  besides   $^{10}$B+d$\to$$\alpha_0$+$^7$Be+n contribute, but  a small background (not larger than $\sim$4\%).} 
\end{center}
\end{figure}

\begin{figure}
\begin{center}
\includegraphics[scale=0.65]{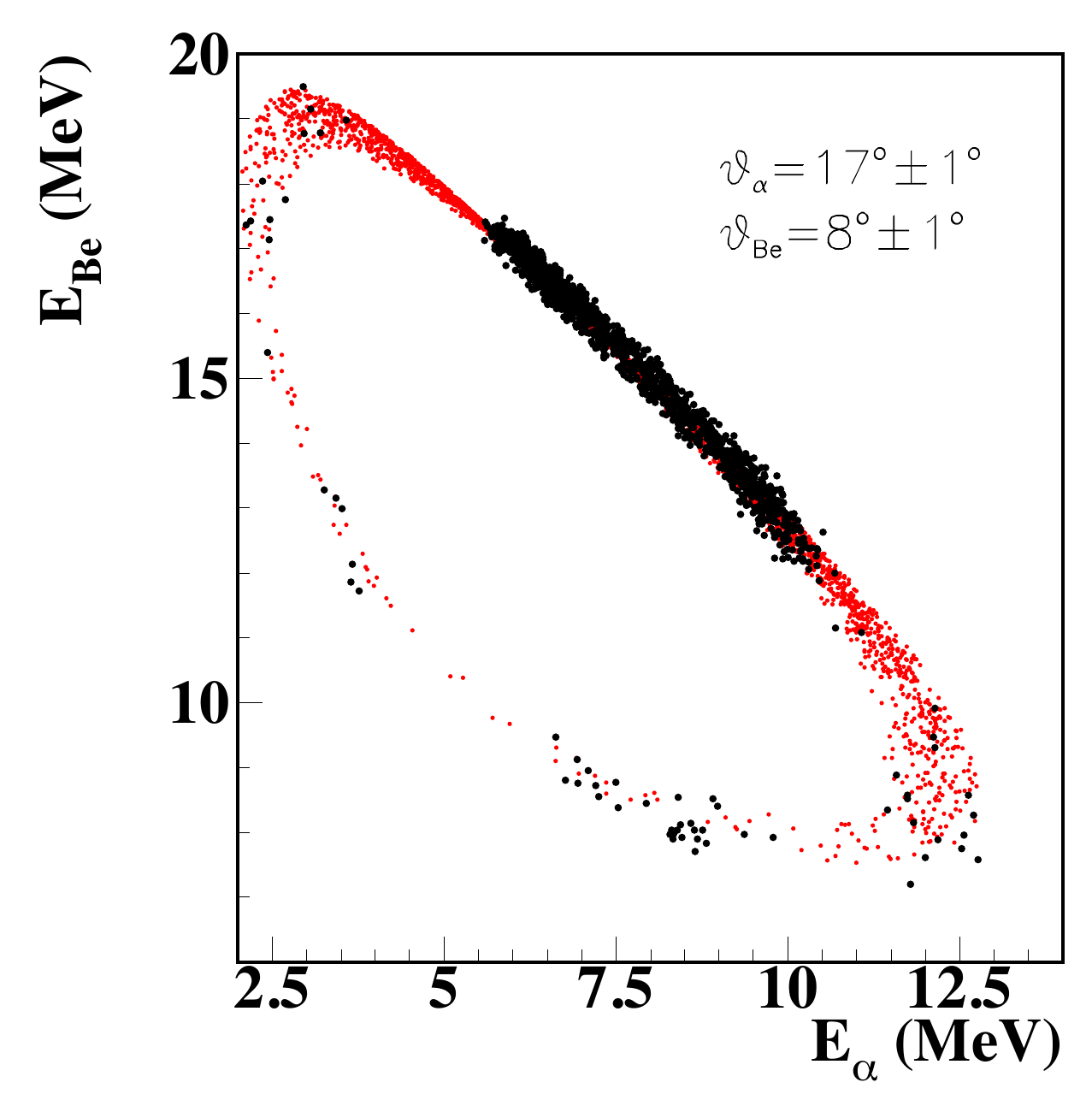}
\caption{(Color online) The experimental kinematical locus E$_{Be}$.vs.E$_{\alpha}$ for the $^2$H($^{10}$B,$\alpha_0$$^7$Be)$n$ reaction (black points) compared with the theoretical one (red points). The comparison has been made for a fixed detection angular pair.}
\end{center}
\end{figure}

\subsection{Detector calibration}
At the initial stage of measurement, masks with 18 equally spaced slits were placed in front of each PSD to perform position calibration.  A correspondence between position signals from PSD's and detection angle  was then established. Energy and angular calibration were performed by using a 9 MeV $^6$Li beam impinging on a CD$_2$ target, to measure reactions on $^{12}$C and $^2$H, and a gold target to measure the $^6$Li+$^{197}$Au elastic scattering. In addition, a three-peaks alpha source ($^{239}$Pu, $^{241}$Am, $^{244}$Cm) was also used for low-energy calibration. The overall procedure lead to a resolution better than 1\% for energy calibration and better than 0.2$^{\circ}$ for angular calibration. 

\section{DATA ANALYSIS}
As already mentioned above, the application of THM requires several steps in the data analysis. Its application is not straightforward and careful evaluation reaction channel and reaction mechanism selections need to be performed. Each of these steps is described in detail in the following paragraphs together with validity tests of the method. 

\subsection{Selection of the  $^2$H($^{10}$B,$\alpha_0$$^7$Be)n channel}
To disentangle the contribution of the $^2$H($^{10}$B,$\alpha_0$$^7$Be)n reaction, $^7$Be nuclei were selected using the standard $\Delta$E-E technique (Fig.3), while no identification was used for $\alpha$ particles on PSD$_B$ and PSD$_C$. In Fig.3  the typical $\Delta$E-E two-dimensional-plot is shown. The kinematical variables have been then reconstructed under the assumption that the mass of the third undetected particle is one (neutron mass).\\Therefore the experimental Q-value spectrum, shown in Fig.4, has been deduced and it is  centered at  about -1.07 MeV, in good agreement with the theoretical value of -1.079 MeV. In the further analysis, only events inside the Q-value peak are considered, being the measured background of Fig.4 lower than the 4$\%$. 
In addition, the experimental $E_{^7Be}$-$E_{\alpha}$ kinematical locus of the $^2$H($^{10}$B,$\alpha_0$$^7$Be)n  reaction was reconstructed and  compared with the simulated one, angle by angle. In particular, Fig. 5 shows the spectra obtained by selecting the angular condition $\theta_{\alpha}$=17$^{\circ}$$\pm$1$^{\circ}$ and $\theta_{Be}$=8$^{\circ}$$\pm$1$^{\circ}$.
Good agreement between the experimental (black solid dots) and theoretical (red solid dots) kinematic loci is found for all the angular couples, the differences in the population of the kinematic loci being originated by reaction dynamics. This procedure confirms the correct identification of the $^2$H($^{10}$B,$\alpha_0$$^7$Be)n reaction channel and the accuracy of the detectors calibration.  
 \begin{figure*}
\begin{center}
\includegraphics[scale=0.6]{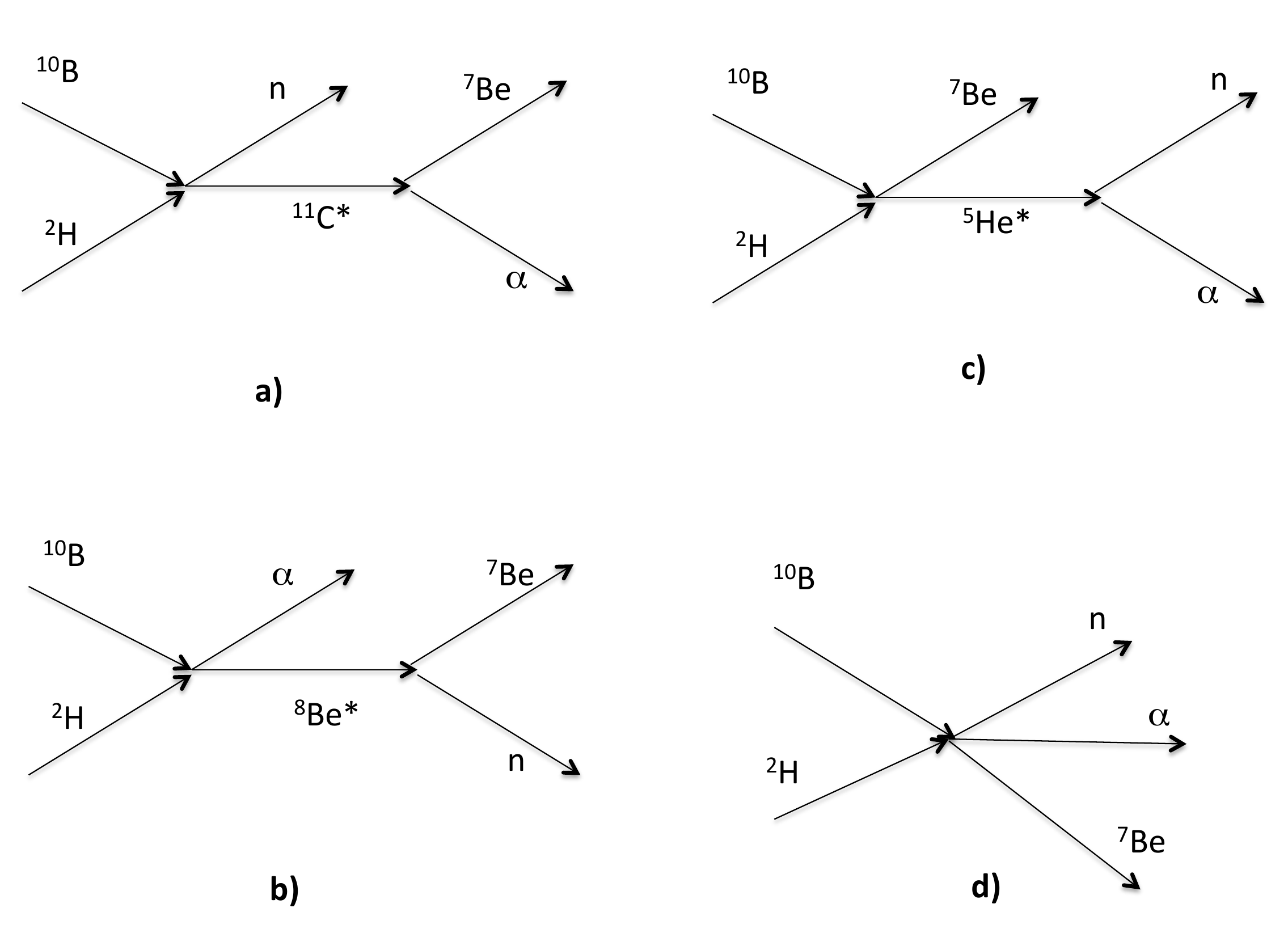}
\caption{Possible simplified diagrams for the $^2$H($^{10}$B,$\alpha_0$$^7$Be)n. Diagrams a), b), c) represent two step processes, proceeding through the formation of the  compound nuclei  $^{11}$C,$^8$Be and $^5$He, respectively. Diagram d) represents a direct breakup mechanism.}
\end{center}
\end{figure*}
\begin{figure}
\begin{center}
\includegraphics[scale=0.65]{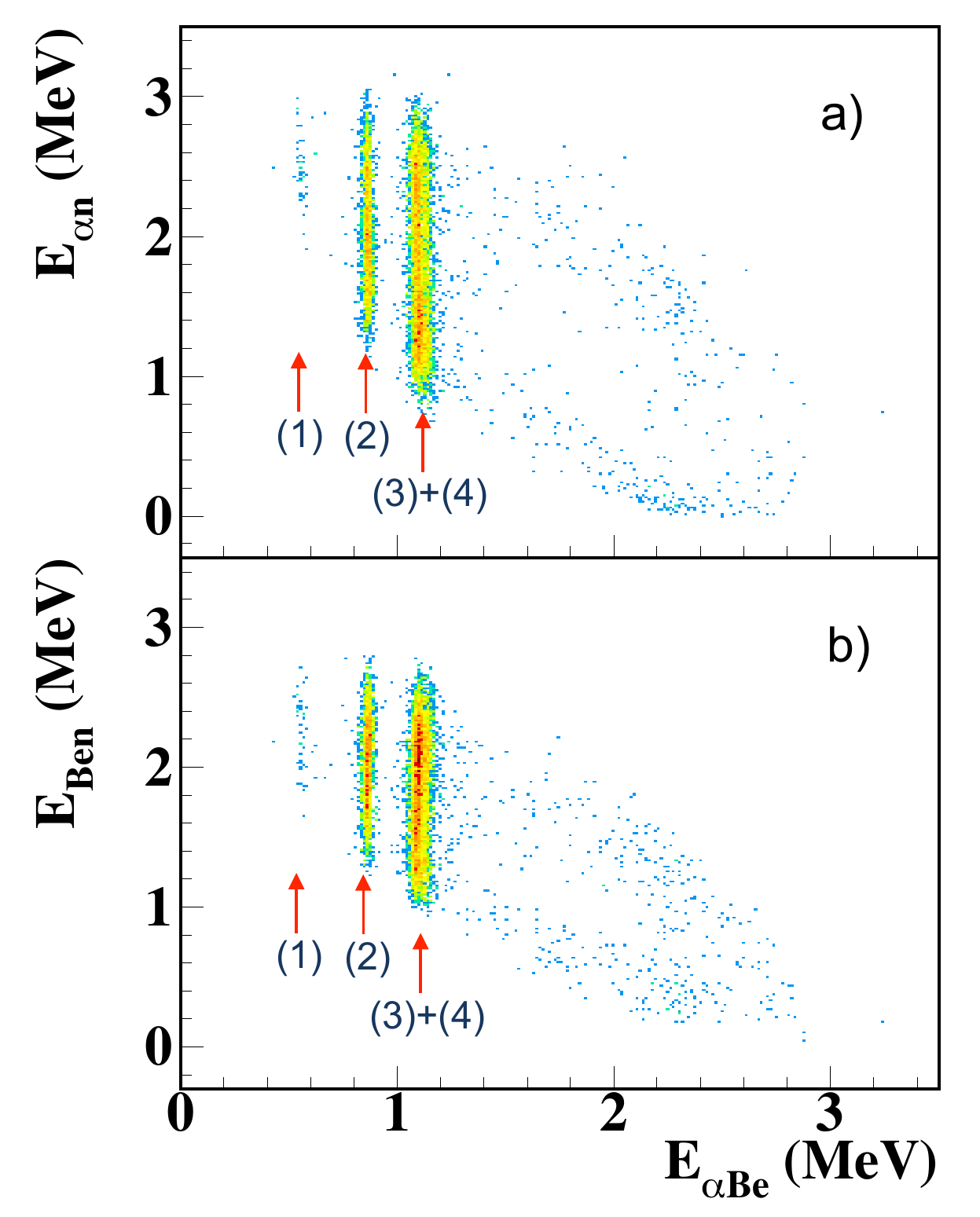}
\caption{(Color online) Two-dimensional plots of the E$_{\alpha-n}$ and E$_{Be-n}$ relative  energies as a function of E$_{\alpha-Be}$. The arrows mark the positions of the 8.104 MeV (1), 8.420  MeV (2) and 8.654 MeV and 8.699 MeV unresolved levels (3)+(4) in $^{11}$C. No evidence of horizontal loci, due to the population of $^5$He and $^8$Be excited levels, respectively, is present.} 
\end{center}
\end{figure}
\subsection{Selection  of the QF reaction mechanism contribution }
The identification of the different reaction mechanisms is a crucial step  in the data analysis because there might be mechanisms other than the QF one, such as sequential decay ($SD$) or direct breakup ($DBU$),  producing the same particles $\alpha$, $^{7}$Be and neutron in the final state (Fig.6). This exit channel can be populated by three different sequential processes, corresponding to the different couplings of the three particles in the exit channel (Fig.6):
\begin{enumerate}
\item  $^{10}$B+$^2$H$\rightarrow$$^{11}$C$^*$+$n$$\rightarrow$$^{7}$Be  + $\alpha$ + $n$ 
\item $^{10}$B+$^2$H$\rightarrow$$^{8}$Be$^*$+$\alpha$$\rightarrow$$^{7}$Be  + $n$ + $\alpha$
\item $^{10}$B+$^2$H$\rightarrow$$^{5}$He$^*$+$^7$Be$\rightarrow$$\alpha$  + $n$ + $^{7}$Be 
\end{enumerate}
Kinematic conditions can be chosen to minimize $SD$ contributions in most cases, as it is possible to identify contributions coming from $SD$ by means of the analysis of the relative energy spectra for any pair of detected particles. 
Fig.7 a) and Fig.7 b)  show the scatter plots of the $^7Be-n$ and $\alpha-n$  relative energies as a function of the $\alpha$-$^7Be$ one. In these plots, any event correlation appearing as a horizontal, vertical or bent line, gives evidence of the formation of an excited intermediate system, finally  feeding the exit channel of interest.\\
The 2D plots of Fig.7 show very clear vertical loci corresponding to $^{11}$C  levels at excitation energies of 8.104 MeV (labelled as (1)), 8.420 MeV (labelled as (2)), 8.654 MeV and 8.699 MeV (unresolved levels labelled as (3)+(4)).  No horizontal loci, corresponding to $^5$He or $^8$Be excited states, are present. Moreover, to determine the presence of the different processes a)-d) of Fig.6, a quantitative analysis has been performed by following the the same approach discussed in several works on QF-mechanisms (see \cite{lattuadanc82, lattuadanc82bis, lattuadanc82tris, calvinc83}) and THM measurements (see \cite{Lacognata2005, Gulino2012, Lamia2012}). In particular, for fixed angles, we have obtained the experimental spectra of the E$_{Be}$, E$_{\alpha}$, E$_{\alpha-Be}$, E$_{\alpha-n}$ kinematical variables. They were compared with Monte Carlo simulations including all the processes of Fig. 6  that can contribute to the reaction yield. The relative weight of each process has been adjusted in order to reproduce experimental data. This analysis leads to a 4$\%$ maximum contribution of process (d) to the total reaction yield and demonstrates that the dominant contribution is given by diagram (a) in Fig. 6.\\ 
Therefore, the $^2$H($^{10}$B,$\alpha_0$$^7$Be)n reaction mainly proceeds through formation of an intermediate $^{11}$C excited nucleus.  In particular, only the 8.699 MeV $^{11}$C excited state  can contribute within the astrophysical  energy region, because  the other three $^{11}$C states at  8.654 MeV, 8.420 MeV and 8.104 MeV are below the $^{10}$B+p decay threshold \cite{Kelley2012}.

\subsubsection{Experimental momentum distribution in PWIA}
A standard way  to investigate the reaction mechanisms is the  study of  the experimental momentum distribution $\mid$${\Phi}(\vec{p_n})$$\mid$$^2_{exp.}$ of $^2$H \cite{Spitaleri2004,Spitaleri2011}, being this quantity very sensitive to the reaction mechanism.
The kinematical variables of the undetected neutron needed to reconstruct  the experimental momentum distribution can be calculated using angles and energies of the detected $\alpha$ and $^7$Be particles.\\
\begin{figure}
\begin{center}
\includegraphics[scale=0.60]{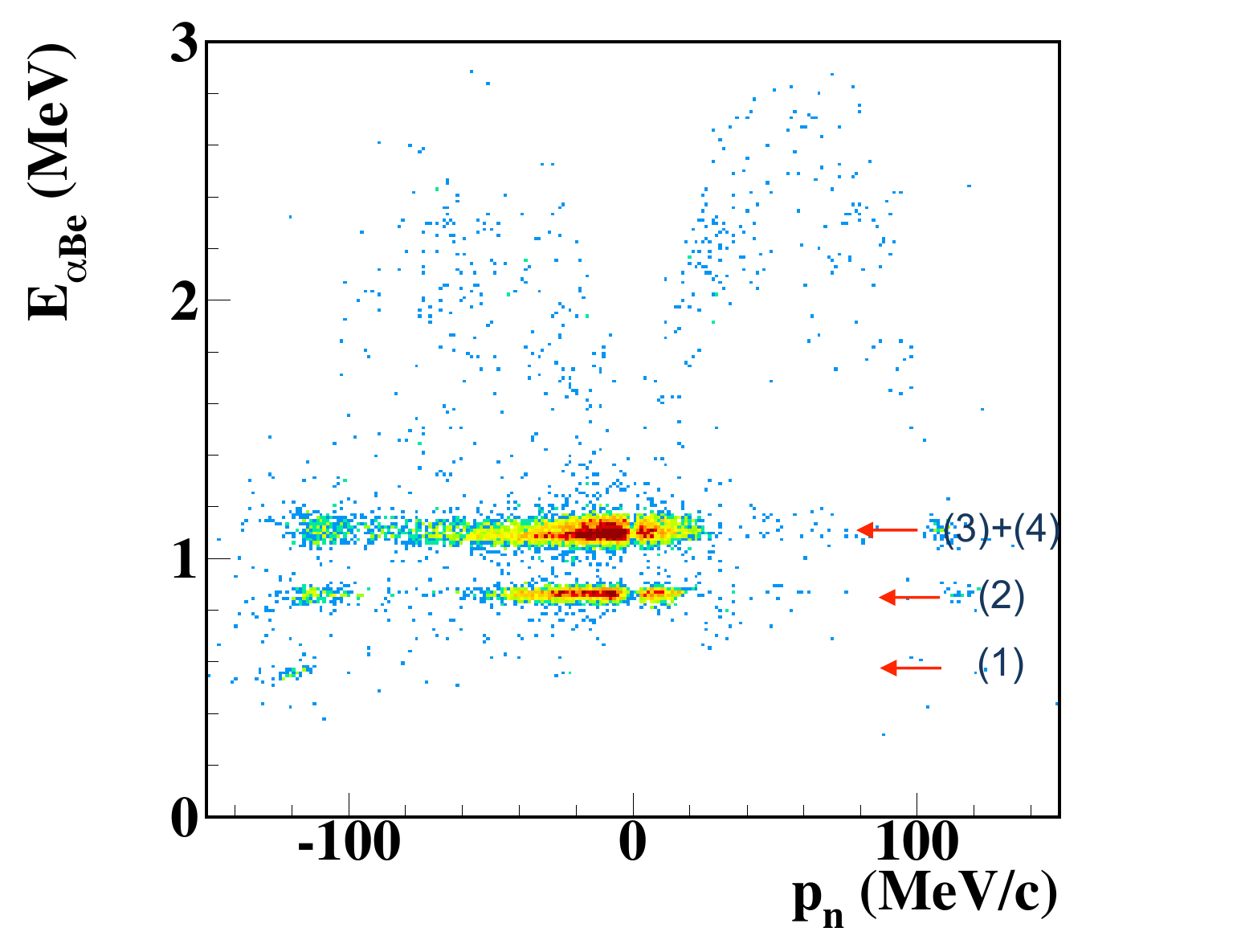}
\caption{(Color online) Two-dimensional plots of the E$_{\alpha-Be}$ relative energy as a function of the experimental momenta values p$_n$ of the undetected neutron. The labels maintain the same meaning as those of Fig.7. It should be noticed that the detected $^{11}$C excited levels populate the low neutron momentum window, corresponding to the kinematical region where the bulk of the QF mechanism is expected.}
\end{center}
\end{figure}
If the factorization of Eq.(3) is applicable, dividing the QF coincidence yield (Y) by the kinematic factor, a quantity which is proportional to the product of the momentum distribution by the p+$^{10}$B $\to$ $\alpha_0$ + $^7$Be two-body cross section is obtained. In a restricted relative energy $\Delta E_{CM}$ and center-of-mass angular range $\Delta \theta_{CM}$, the differential binary cross section $\frac{d\sigma}{d\Omega}$ can be considered almost constant  and from Eq.(3) we obtain the simple relation:
\begin{equation}
\mid{\Phi}(\vec{p_n})\mid^2_{exp.} \propto \frac{Y}{KF}
\end{equation}
The experimental  momentum distribution $\mid$${\Phi}(\vec{p_n})$$\mid$$^2_{exp.}$ has been obtained by following the standard approach given in \cite{Spitaleri2004}, by considering the 2D-plot  $E_{\alpha-Be}$.vs.p$_n$ shown in Fig.8. By selecting the $E_{\alpha-Be}$ events corresponding to a very narrow window in both relative energies and angles, a projection onto the p$_n$ axis has been made giving the experimental yield Y used in the previous formula.\\ 
Neutron momentum values ranging from -100 MeV/c to 100 MeV/c were deduced, accordingly to the horizontal axis of Fig. 8. These data were then corrected for the kinematical factor, thus removing phase-space effects. Finally, an average between the experimental yield corresponding to the condition -100 MeV/c$<$p$_n$$<$0 MeV/c and the one corresponding to the condition 0$<$p$_n$$<$100 MeV/c has been performed.\\ The resulting momentum distribution is given as black symbols in Fig. 9, as a function of the modulus of the neutron momentum $\mid$$\vec{p_n}$$\mid$.
It represents the experimental momentum distribution as deduced from the present $^2$H($^{10}$B,$\alpha_0$ $^7$Be)n measurement performed at E$_{beam}$= 24.5 MeV.
The black solid  line in Fig.9  is the theoretical distribution given by the squared Hulth\'en wave function in momentum space:  
\begin{eqnarray}
\mid{\Phi}(\vec{p_n})\mid^2=  \frac{1}{\pi} \sqrt{\frac {ab(a+b)}{(a-b)^2}}\left[\frac{1}{a^2+  p_n^2}-\frac{1}{b^2+ p_n^2}\right]
\end{eqnarray}		
normalized to the experimental maximum,  with  parameters  $a$=0.2317~fm$^{-1}$ and $b$=1.202~fm$^{-1}$ \cite{Zadro1989}. The experimental full width obtained in the present work is 54$\pm$ 5 MeV/c. 
\begin{figure}
\begin{center}
\includegraphics[scale=0.50]{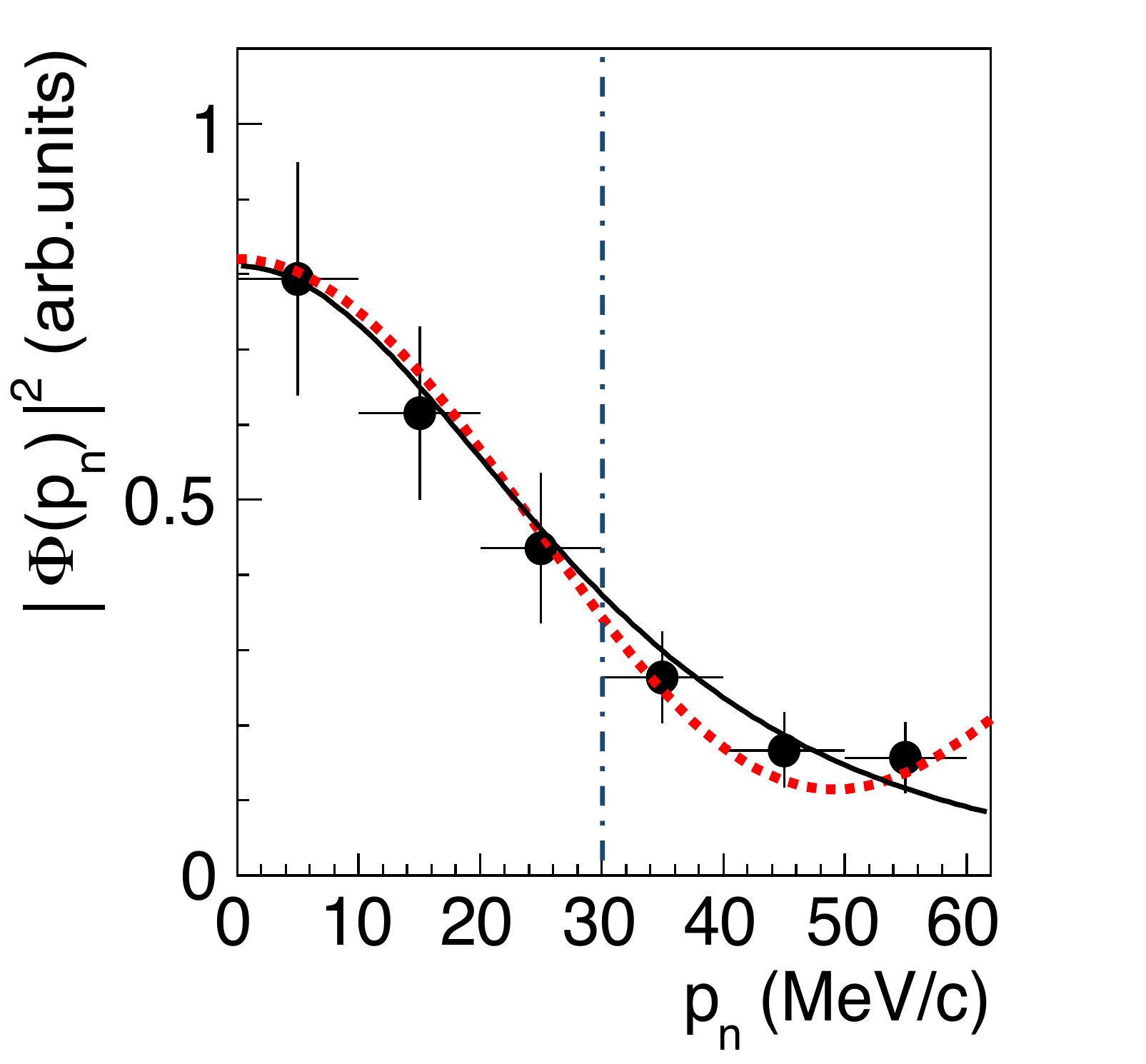}
\caption{(Color online) Experimental momentum distribution (black points) compared with the theoretical one given by the squared Hulth\'en wave function in momentum space (black line) and the one given in terms of a DWBA calculation performed via the FRESCO code (red dashed line). The error bars include only statistical errors. The vertical blue line delimits the momentum region $p_n$$\leq$30MeV/c selected for the further analysis.}
\end{center}
\end{figure}
\subsubsection{Comparison between the PWIA and the DWBA calculations}
The PWIA framework is usually adopted in the THM application since it accurately describes the experimental data, provided that 
the appropriate FWHM (full width at half maximum) for the experimental value of the momentum transfer is introduced into the calculations \cite{Pizzone2005,Pizzone2011}. This is simply accounted for by using the experimental momentum distribution to extract the HOES cross section. The validity of a PWIA approach can be verified employing the Distorted Wave Born Approximation (DWBA). For such a reason, a DWBA calculation has been additionally performed by means of the FRESCO code \cite{thom}, by considering the optical model potential parameters given in Perey and Perey \cite{ Perey76}. The result is shown as dashed red line in Fig.9, after normalization to the experimental data. From the comparison with the experimental momentum distribution one can state that, if we limit our event selection to the region close to the maximum of the experimental momentum distribution ($p_n$=0 MeV/c for $s-wave$ relative motion), the DWBA approach and  the PWIA one give similar results, apart from an inessential scaling factor.  In fact, the THM cross section is expressed in arbitrary units.
The momentum distributions in PWIA (black solid line) and DWBA (red dotted line) nicely agree with the experimental data over the whole neutron momentum range given by Eq. (7) \cite{Shapiro1967}. However, to select only the experimental data for which the contribution of  the QF reaction mechanism is dominant and the differences between PWIA and DWBA are negligibly small, the narrower 0-30 MeV/c momentum range (delimited by the vertical dot-dashed line in Fig.9) was chosen for the next analysis.
\subsection{Selection of the events for the $^{10}$B(p,$\alpha$)$^7$Be investigation}
The selected events are finally shown in the two panels of Fig. 10 as a function of $^{11}$C excitation energy. In particular, the upper panel shows the well separated peak at about 8420 keV, while in the lower panel the convolution between the 8654 keV and the 8699 keV levels is reported.
\begin{figure}
\begin{center}
\includegraphics[scale=0.65]{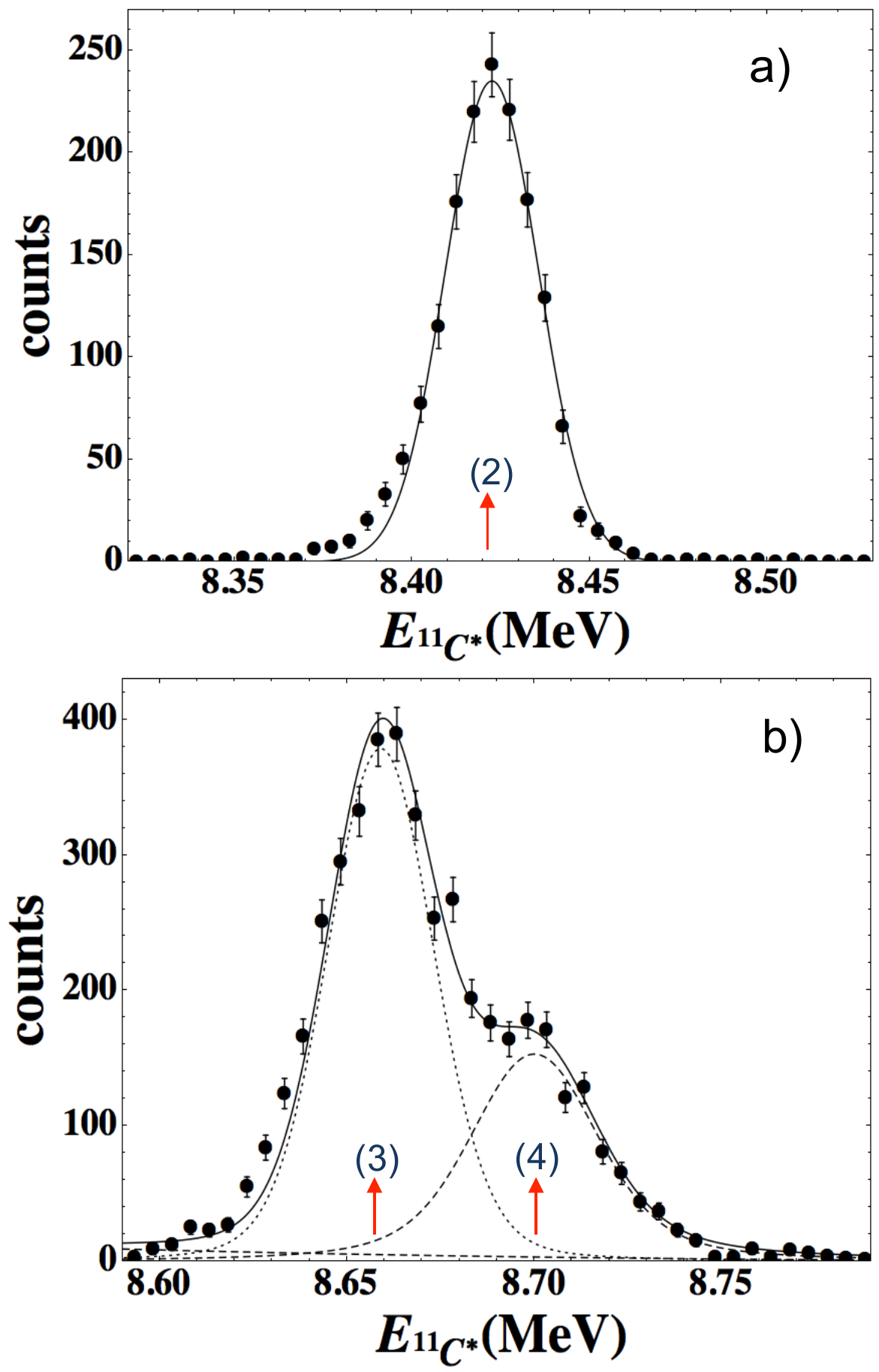}
\caption{(Color online) Events corresponding to the kinematical condition 0$\leq$p$_n$$\leq$30 MeV/c (as discussed in the text). Panel a) shows the events corresponding to the 8.420 MeV $^{11}$C level (2), while in panel b) the events corresponding to the two unresolved 8.654 MeV and 8.699 MeV levels (3)+(4) in $^{11}$C are displayed.}
\end{center}
\end{figure}
%
\begin{table}[h]
\caption{Resonance energies of excited $^{11}$C states (E$^*$), the corresponding  E$_{cm}$ in the $^{10}$B-p system, the natural width $\Gamma_{cm}$ (from literature), and the experimental width $\Gamma_t$ obtained in this work.}
\begin{tabular}{lllllll}
\hline
\hline   
 {}&E$^*$                     {}&$[E_{cm}]$          {}&J$^\pi$             {}& $\Gamma_{cm}$              {}&$\Gamma_t$                       {}&Ref.     \\ [1.5 ex]
 {}&(keV )                   	                      {}& (keV)                   {}&                         {}&(keV)                                   {}&(keV)		                                                     \\[1.5 ex]
\hline
  {}&8104$\pm$1.7                               {}&-580                     {}&3/2$^-$        {}&6$^{+12}_{-2}$ $\cdot$10$^{-3}$     {}&---             {}&\cite{Kelley2012}                           \\[1ex]
  {}&8420$\pm$2                               {}&-287                     {}&5/2$^-$        {}&8$\cdot$10$^{-3}$     {}&31$\pm$3                       {}&\cite{Kelley2012}                 \\[1ex]
  {}&8654$\pm$4                                {}&-35                        {}&7/2 $^+$     {}&$\leq$5                             {}&34$\pm$2                 {}&\cite{Kelley2012}                    \\[1ex]
  {}&8699$\pm$2                                {}&10                        {}&5/2$^+$        {}&16$\pm$1                          {}&40$\pm$2                {}&\cite{Wiescher1983}                       \\[1ex]
\hline
\hline
\end{tabular}
\end{table}
The isolated 8.420 MeV level has been fitted with a Breit-Wigner function, giving the following parameters:  resonance energy  E$_R$ = 8.422$\pm$ 0.002 MeV, $\sigma$=13$\pm$1 keV and FWHM $ \simeq$31$\pm$3 keV. These must be compared with those in Table II, $\Gamma$$\sim$8 eV and E$_R$ = 8.420 MeV as given in the literature \cite{Kelley2012}.\\
Since  the isolated level of Fig.10 a) is very narrow, we can conclude that the total energy resolution is equal to its  experimental width  $\Delta E_{res.}$=31$\pm$3 keV (FWHM) and it is assumed to be constant over  the whole measured energy range. The levels labelled with $(3)$ and $(4)$ in Fig.10 b)  correspond  to the unresolved 8.654 MeV and 8.699 MeV $^{11}$C excited states, whose overlap is due to the experimental energy resolution. To select events corresponding to the region with energy $E_{cm}$$\geq$0,  it is necessary to separate this two contributions and to evaluate the uncertainties coming from such a procedure.\\
Since the resonance  energy $E_R$ and the width  $\Gamma_i$ of these two unresolved resonances are known \cite{Kelley2012}, the observed peaks of Fig.10 b) have been fitted by considering the broadening by energy resolution effects, previously described, on the function F(E)$_{unres.}$. This function is expressed in terms of the incoherent sum of two Breit-Wigner shapes $bw(E)_{(3)}$ and $bw(E)_{(4)}$  plus a non resonant contribution $p(E)$:
\begin{equation}
F(E)_{unres.}= bw(E)_{(3)} +  bw(E)_{(4)} + p(E) 
\end {equation}
where 
\begin{eqnarray} 
bw(E)_{(i)} = N(E_{R_{(i)}})\cdot \frac {\left(\frac{\Gamma_{(i)}}{2}\right)^2}{\left(E-(E_{R_{(i)}}\right)^2 + {\left(\frac{\Gamma_{(i)}}{2}\right)^2}}
\end{eqnarray}
where the parameters of Eq.(11) are:\\
- E$_{R_3}$=8.654 MeV the energy resonance (3),\\
- N(E$_{R_3}$)=1830$\pm$48  the peak value in correspondence of resonance (3),\\
-  $\Gamma_3$=5 keV   the width of resonance (3),\\  
- E$_{R_4}$=8.699 MeV the energy resonance  (4),\\ 
- N(E$_{R_4}$)=306$\pm$18 the peak value in correspondence of resonance (4),\\
- $\Gamma_4$=16 keV  the width of resonance  (4)\\ 
and
\begin{eqnarray}
 p(E) =  3.75 -39.24\cdot(E - E_{thr.}) +\nonumber\\ 
  + 143.25\cdot(E-E_{thr.})^2  -168.33\cdot(E-E_{thr.})^3.
\end{eqnarray}
being E$_{thr.}$=8.689 MeV the proton decay threshold for the $^{11}$C nucleus.
The procedure described above returns the full-black line superimposed on the TH data of Fig.10 b), giving a reduced  $\chi^2$ of $\sim$1.7.\par
Because of the presence of the subthreshold 8.654 MeV level, its contribution has been properly subtracted for the experimental data of Fig.10 b) lying in the window 0$\leq [E_{cm}]_i\leq$100 keV. \\
The corresponding uncertainty ($\epsilon_{lev.sub.})_i$ has been then evaluated as
\begin{eqnarray}
 (\epsilon_{lev.sub})_i= \frac{N_{ev}(E_i)^{[(3)+(4)]}-N_{ev}(E_i)^{(3)}}{N_{ev}(E_i)^{[(3)+(4)]}}
\end{eqnarray}
where  $N_{ev}(E_i)^{[(3)+(4)]}$ and $N_{ev}(E_i)^{(4)}$ are the number of events corresponding to $F(E_i)_{unres.}$  and to $bw(E_i)_{(2)}$ at the  energy $E_i$,  respectively.\\
In Fig.10 b),  the   fit of the unresolved levels (3)+(4) (solid line)  is shown as well as  the separate  level contributions (dotted (3) and dashed (4) lines).The contribution of the 8.699 MeV $^{11}$C excited  level separated from the subthreshold 8.654 MeV state is shown in Fig.11. Note that the solid line in Fig.11, corresponding to the fit  reported in  Fig.10 b),  is obtained  by taking in account the resonant  (upper dashed line) and non resonant   (lower dashed line) contributions, while  errors affecting the data points are the statistical only.\\ 
In the next phase of data analysis, only  these events are taken into account for extracting the $^{10}$B(p,$\alpha_0$)$^7$Be S(E)-factor.\\

\begin{figure}
\begin{center}
\includegraphics[scale=0.52]{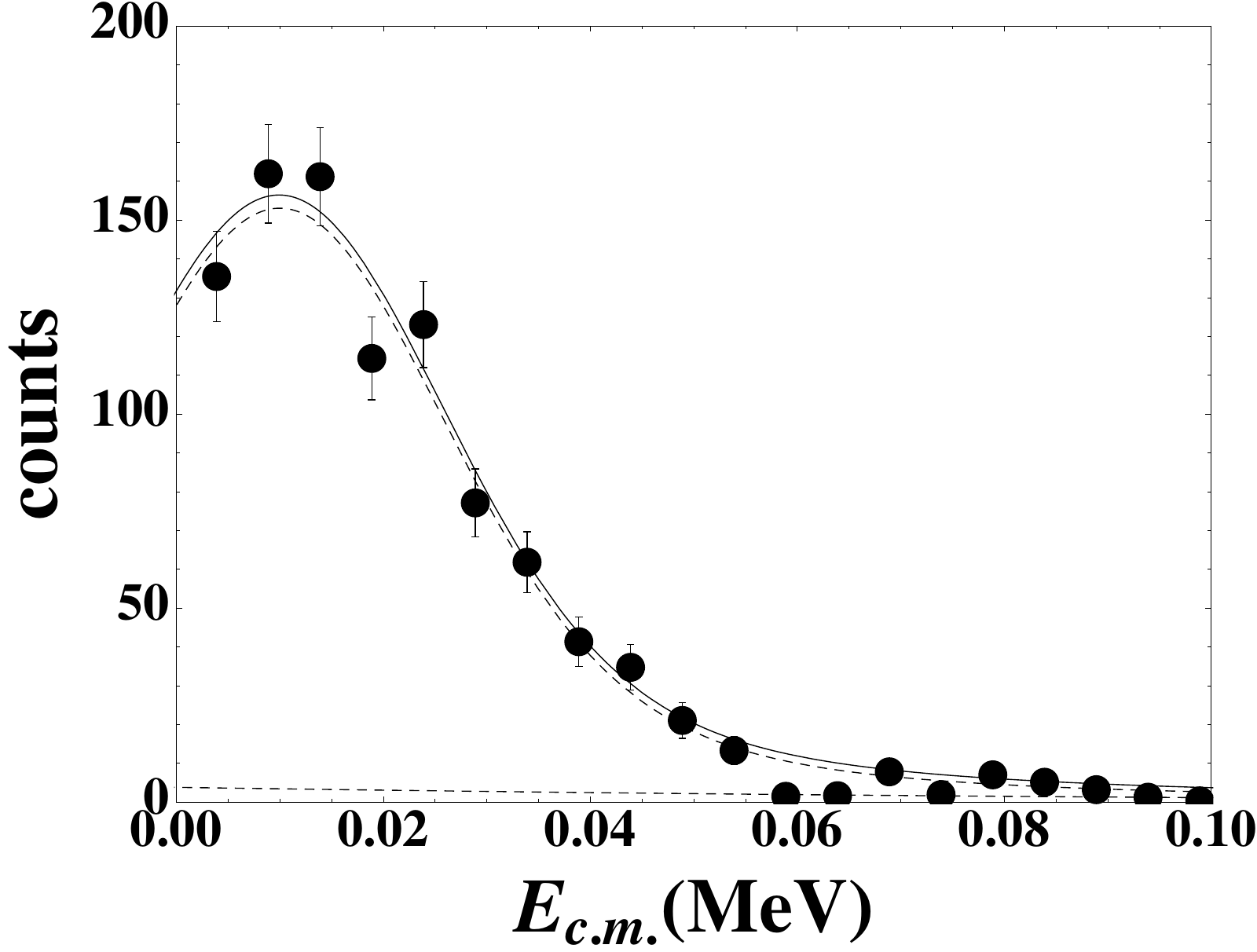}
\caption{Selected events corresponding to the 10 keV resonance for the$^{10}$B(p,$\alpha_0$)$^7$Be reaction after removing the sub-threshold contribution due to the 8654 keV $^{11}$C resonant level, as discussed in the text.} 
\end{center}
\end{figure}

\section{RESULTS}

\subsection{Two-body cross section}
The $^{10}$B(p,$\alpha$)$^7$Be HOES differential cross section is extracted by inverting Eq.(3):
\begin{equation}
\left (\frac{d\sigma(E)}{d\Omega}\right)^{HOES}\propto \frac{d^3\sigma}{d\Omega_\alpha d\Omega_{^7Be} dE_\alpha} \cdot ({KF\cdot \mid{\Phi}(\vec{p_n})\mid^2_{exp.}})^{-1}\\
\end{equation}
The product $KF\cdot $ $\mid$${\Phi}(\vec{p_n})$$\mid$$^2_{exp.}$ is  calculated by using  a Monte Carlo simulation, including masses, angles and momenta of the detected $^7$Be and alpha particles, and the experimental momentum distribution obtained above.\\ 
As already mentioned, since the proton is brought inside the $^{10}$B nuclear field, the binary reaction is HOES and  represents only the nuclear part \cite{Spitaleri2011,Spitaleri2011b,Spitaleri2004}. For this reason, the effects of the Coulomb barrier must be introduced to compare the differential cross section in to the on-energy-shell one.
The so-called TH cross section is then defined using the relation:
\begin{equation}
\left [\frac{d\sigma(E)}{d\Omega}\right]^{TH} =  \left [\frac{d\sigma(E)}{d\Omega}\right]^{HOES} \cdot  P_0(k r)
\end{equation}
where the penetration probability  P$_{l=0}(k r)$ = P$_0(k r)$ of the Coulomb barrier  is defined by the equation:
\begin{equation}
P_0(k r) =  \frac{k  r}{F^2_0(k r)+G^2_0(kr)}
\end{equation}
with $F_0$ and $G_0$  regular and irregular Coulomb  functions for  $l = 0$,  $ k$ and $r$  the  relative wave number and  the interaction radius for the  $p- ^{10}$B system, respectively.
 Since the angular distributions for the  $^{10}$B(p,$\alpha_0$)$^7$Be reaction  are almost isotropic    \cite{Youn1991}, the differential  cross section integrated over the experimental $\theta_{cm}$ range differs from the total cross section  $\sigma(E)$ by an inessential scaling factor. \par
In the case of the   $^{10}$B(p,$\alpha_0$)$^7$Be reaction,  the $l$=0 contribution is dominant  as the $^{10}B$ ground state has $J^\pi $= 3$^+$,  the proton has $J^\pi$ = 1 / 2$^+$  and the  8.699 MeV level $J^\pi$ = 5/ 2$^+$.
The small non-resonant background is  represented by an  $l= 0$ component in the region of astrophysical interest, thus  the bare nucleus  total cross section can be calculated by using:
\begin{eqnarray}
\sigma(E) = W_0 \cdot  {P_{0}(k r)}\cdot [\sigma(E)]^{HOES}= \nonumber\\
 = W_0\cdot [\sigma(E)]^{TH}
\end{eqnarray}
where W$_0$ is a normalization constant to be determined.

\subsection{Bare nucleus astrophysical S$_b$(E)-factor}
The determination of the bare nucleus THM S$_b$(E)-factor in absolute units has been then performed by using the available direct data of \cite{Angulo1993, Youn1991,Angulo1999}, showed in Fig.12. However, low-energy direct measurements are strongly affected by the electron screening effects \cite{Assenbaum1987, Strieder2001}, thus the absolute scale on the S$_b$(E)-factor needs to be obtained by normalizing the TH data to the OES one in an energy range where the electron screening effects are negligible to reduce systematic errors. In addition, energy resolution effects alter the energy trend of the present TH data, thus the normalization procedure is not straightfoward.\\
\begin{figure}
\begin{center}
\includegraphics[scale=0.50]{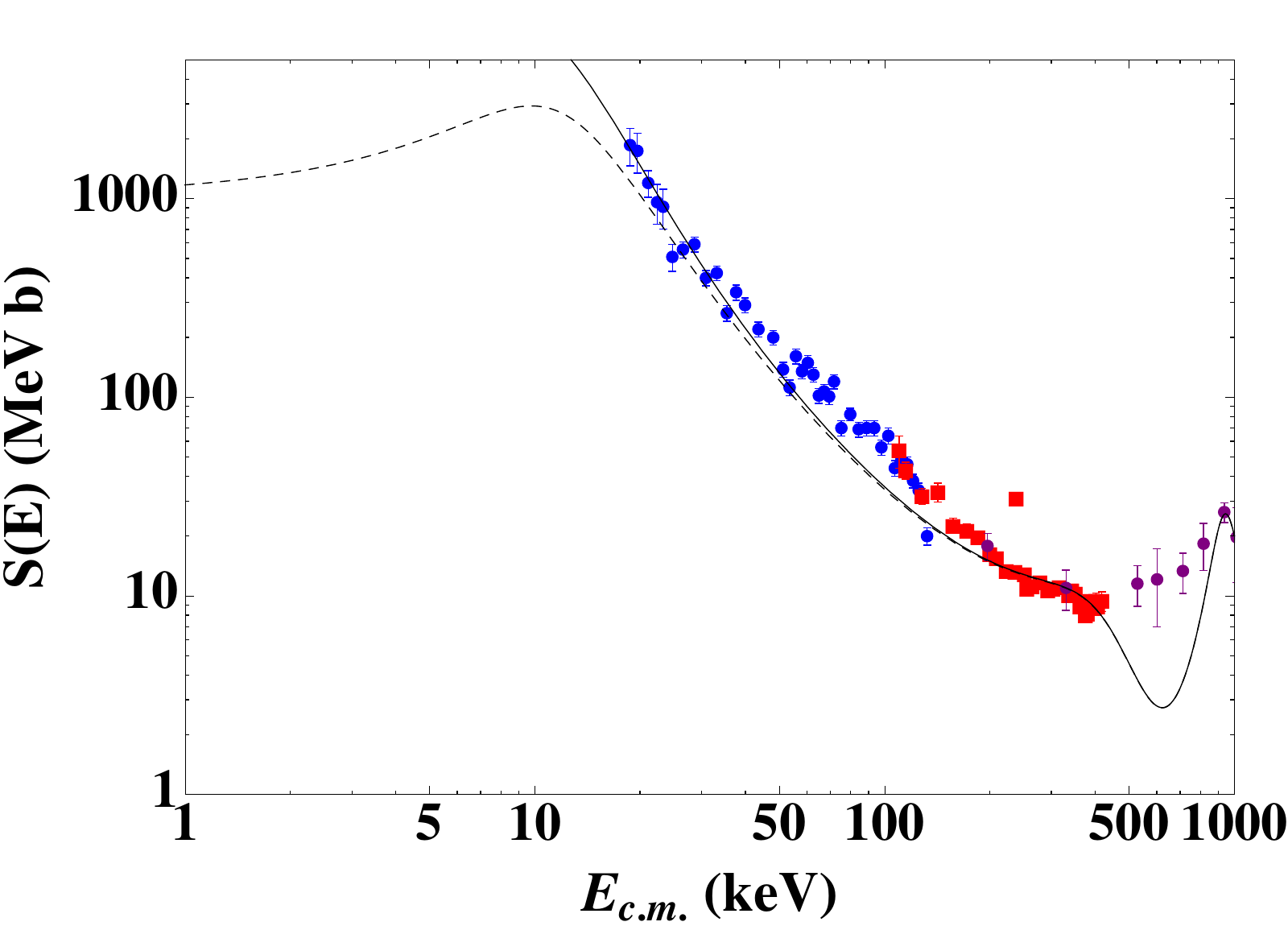}
\caption{(Color online) Direct astrophysical S(E)-$factor$ of the  $^{10}$B(p,$\alpha_0$)$^7$Be reaction \cite{Angulo1993, Youn1991,Angulo1999}. The lines represent the $R$-matrix calculation with the resonance parameters from the literature \cite{Wiescher1983, Kelley2012}, for bare (dashed line) and screened (full line) nuclei. Red symbols are used to mark the data of \cite{Youn1991}, corrected for the factor 1.83 as done in \cite{Angulo1993}, blue symbols refer to the measurement performed by \cite{Angulo1993}, and purple symbols refer to the thick-target measurements of \cite{atdata79}. All these data are included in the NACRE compilation of \cite{Angulo1999}.}
\end{center}
\end{figure}
For such a reason, a function describing the available direct S-factor measurements was then deduced and reduced to the same experimental resolution of the THM S$_b$(E)-factor, thus allowing finally to get the normalization coefficient.\\
The available low-energy direct data from Refs. \cite{Youn1991,Angulo1993, Angulo1999} have been described by means of an $R$-matrix calculation, performed by using the parameters of the relevant resonances currently reported in literature \cite{Wiescher1983, Kelley2012} (Table III). The enhancement at energies lower than 50 keV has been described by using the electron screening potential value of 430 eV given in \cite{Angulo1993}.\\
\begin{table}[htdp]
\caption{The resonance parameters used in the $R$-matrix calculation, as given in the literature \cite{Wiescher1983, Kelley2012}.}
\begin{center} 
\begin{tabular}{lllllllr}
\hline 
\hline \\
{}&E$_r$       {}&$\Gamma_p$   {}&$\Gamma_{alpha}$   {}&$\Gamma^{Fit}_{Tot}$     {}&$\Gamma^{Lit}_{Tot}$         \\ [1.3ex] 
{}&(keV)            {}&(keV)              {}&(keV)                      {}&(keV)                              {}&(keV)                                         \\ [1.3ex] 
\hline \\ 
{}&9.4               {}&2$\cdot$10$^{-17}$ 	{}&15        {}&15                                   {}&15		                         \\ [1.0ex]      
{}&500              {}&3.3$\cdot$10$^{-4}$       {}&500      {}&500                                     {}&500                                  \\ [1.0ex]
{}&945              {}&---       {}&---      {}&210                                     {}&210                                  \\ [1.0ex]           
 \hline 
\end{tabular}
\end{center}
\end{table}
Fig.12 shows the available direct S(E)-factor measurements data  for the  $^{10}$B(p,$\alpha_0$)$^7$Be reaction as reported in the literature (red and blue symbols for \cite{Youn1991,Angulo1993, Angulo1999}, respectively) and the obtained $R$-matrix calculation (solid line).\\
The very poor reduced $\chi^2$ ($\chi^2$$\sim$8) urges to perform new improved direct measurements of the $^{10}$B(p,$\alpha_0$)$^7$Be reaction. Indeed, the $R$-matrix calculation nicely describes the astrophysical factor at about 500 keV and below about 50 keV, while it fails to reproduce the astrophysical factor in the energy region where the two direct data sets overlap, suggesting the presence of some systematic effect. The R-matrix calculation in Fig.12 includes the additional 9645 keV $^{11}$C level, determining a resonance at about 945 keV in the $^{10}$B-p center-of-mass system. For this resonance, the reduced widths were chosen to supply the $\Gamma$ in the literature \cite{Kelley2012}. However, a strong disagreement is evident between the R-matrix calculation and the experimental direct data reported in \cite{atdata79}. A possible explanation is that these data were deduced using the thick-target approach for which no proper deconvolution procedure was operated by the authors.\\  
Finally, it must be noticed that the $R$-matrix calculation has been also performed by considering on the data of \cite{Youn1991} the same correction factor used by \cite{Angulo1993}. In this sense, new direct measurements at higher energies could be used to constrain such a fit, for both absolute values and adopted resonance parameters.\\
The THM  S$_b$(E)-factor of Fig.13 (black dots) has been then obtained by normalizing it to the $R$-matrix calculation of Fig.12 smeared to match the same experimental resolution of the present experiment. The normalization procedure, performed in the energy range 50-100 keV in which electron screening does not strongly alter the pure resonant trend of the astrophysical S(E)-factor and in which the resonant 945 keV level does not play any significant role (less than 2\%), leads to an overall uncertainty of about 15\%  with a reduced  $\chi^2$  of 0.5.\\ 
\begin{figure}
\begin{center}
\includegraphics[scale=0.44]{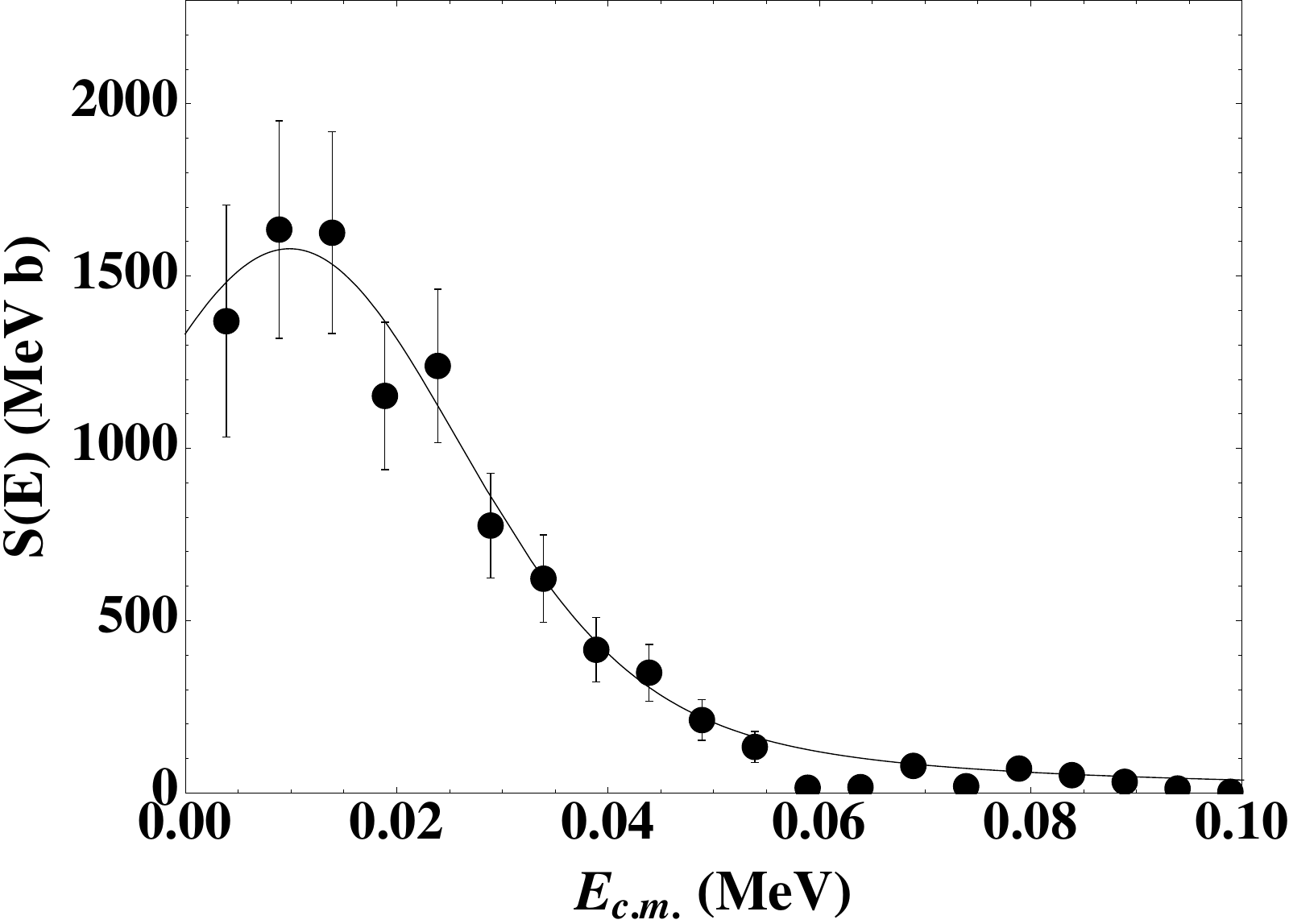}
\caption{The experimental TH S(E)-factor (black dots) together with its fit (solid line). The error bars include the sources of uncertainty described in the text.}
\end{center}
\end{figure}
The error bars of Fig.13 include the statistical error,  the uncertainty connected to the sub-threshold level subtraction, the uncertainty derived from  the choice of the nuclear  radius in the penetrability factor ($r_0$ in $P_0$,  Eqs.15,16), and the uncertainty due to the normalization procedure. Table IV lists the values of the THM S(E)-factor together with the total uncertainty.\\
A fit to the data was performed to evaluate the S$_b$(E)-factor at zero relative energy. Since this fit has the sole aim to obtain such numerical value and not to provide resonance parameters with a physical meaning, a simple functional form has been used, given by the sum of  first order polynomial and  a Gaussian function with parameters:
 \begin{equation}
S (E) = [a_0 + a_1 E]+ N_{E_R} \cdot e^{-\frac{(E- E_R)^2}{ 2\sigma ^2}}
\end{equation}
with $a_0, a_1$, the peak value $N(E_R)$,  the width $\sigma$, and $E_R$ as free parameters.\par
The best fit parameters are  $E_R$ = 0.010$\pm$0.002 MeV,  $N_{E_R}$= 1315$\pm$79 MeV/b,  $\sigma$ = 0.016$\pm$0.002 MeV, $a_0$=236$\pm$59 MeVb, $a_1$= -2320$\pm$614 MeV. A reduced $\chi^2$ of 0.6 is obtained.\\
\begin{table*}[htdp]
\caption{Values of the THM astrophysical  S(E)-factor  at infinite resolution and at the THM energy resolution (31 keV, S(E)$_{31 keV}$), as a function of the E$_{c.m.}$ $^{10}$B-p relative energy. $\Delta$S(E) and $\Delta S(E)_{31 keV}$ are the corresponding uncertainties. The statistical $\epsilon_{stat}$ and the level subtraction $\epsilon_{lev.sub.}$ uncertainties are also reported and, finally, the total percentage error. Additional sources of uncertainties are: the effect of the change on the interaction radius $r_0$ on the penetration factor (2$\%$) and the normalization error (about 15$\%$).}
\begin{center} 
\begin{tabular}{ccccccccccccccc}
\hline
\hline\\
{}&$E_{cm}$         {}&S(E)       {}&$\Delta$S(E)  {}&$S(E)_{31 keV}$       {}&$\Delta$$S(E)_{31 keV}$  {}&$\epsilon$$_{stat.}$   {}&$\epsilon$$_{lev.sub.}$    {}&$\epsilon$$_{tot.}$     \\[+1.5 ex]
{}&(keV)                {}& (MeV b)         {}& (MeV b)   {}& (MeV b)         {}& (MeV b)       			  {}& $\%$.              		   {}&$ \%$	  				   {}&$\%$        		       \\[1.5ex]
\hline 
\\

{}&3.9      	    {}&1995         {}&499   		  {}&1368            {}&342           	{}&9                    {}&17                 {}&25                          \\[1ex]
{}&8.9      	    {}&3071         {}&583   		  {}&1634            {}&310            	 {}&8        		  {}&9                {}&19                           \\[1ex]
{}&13.9    	    {}&2530         {}&455		  {}&1625            {}&292             {}&8          		 {}&6                   {}&18                           \\[1ex]
{}&18.9    	    {}&1411         {}&268		  {}&1151           {}&219             {}&9         		 {}&5                   {}&19                           \\[1ex]
{}&23.9    	    {}&797           {}&143 	       {}&1239            {}&223        	     {}&9                    {}&4                  {}&18                           \\[1ex]
{}&28.9   		    {}&496           {}&99 		  {}&775              {}&155             {}&11                 {}&5	       	      {}&20                        \\[1ex]
{}&33.9   		    {}&336           {}&67	  	       {}&621              {}&124 	      	{}&13	                 {}&5		           {}&20  		                \\[1ex]
{}&38.9   		    {}&244           {}&56		       {}&415              {}&95 		     {}&16	                 {}&6	         	      {}&23 		                \\[1ex]
{}&43.9   		    {}&185           {}&44		       {}&348              {}&83	          {}&17	                 {}&6		           {}&24                          \\[1ex]
{}&48.9  		    {}&146           {}&41		       {}&211             {}&59 		     {}&22	                 {}&9	      		 {}&28                         \\[1ex]
{}&53.9   		    {}&119           {}&39		       {}&132             {}&44 			{}&27                  {}&12		           {}&33		                    \\[1ex]
{}&58.9  		    {}&99           {}&97		       {}&15               {}&15		   	{}&83                   {}&51  			 {}&98		                    \\[1ex]
{}&63.9  		    {}&84            {}&77		       {}&16               {}&15		           {}&78			     {}&46		      {}&92 		                \\[1ex]
{}&68.9  		    {}&72            {}&29  		  {}&78              {}&32			     {}&36		             {}&14	                 {}&41		                     \\[1ex]
{}&73.9                 {}&63             {}&52    	    	  {}&19               {}&16                {}&73                    {}&37          {}&83                          \\[1ex]
{}&78.9   		   {}&56              {}&24 		  {}&70               {}&30		      {}&38			  {}&12   				{}&43 		                \\[1ex]		   
{}&83.9   	 	   {}&49              {}&24    		  {}&51               {}&25		  	 {}&44			  {}&15	           {}&49  		                \\[1ex]
{}&88.9   		   {}&44             {}&27  		  {}&32               {}&19	       	 {}&56			       {}&21			 {}&61  	                     \\[1ex]
{}&93.9   		   {}&40             {}&40	            {}&12               {}&12	   	      {}&90			       {}&38		      {}&99		                     \\[1ex]
{}&98.9   	        {}&36            {}&36         	  {}&3                 {}&3		           {}&100     		  	   {}&71		      {}&100		                \\[1ex]
{}&103.9  	   {}&33             {}&23		       {}&24               {}&17            	  {}&65	 		  {}&22	            {}&71                        \\[1ex]
\hline
\hline
\end{tabular}
\end{center}
\end{table*}
\subsection{Electron screening}
In order to compare the THM data fit with the ones reported in the literature, it has been necessary to remove the effect of the energy resolution affecting the THM data, causing a broadening of the resonant peaks. For such a reason, the TH S(E)-factor at infinite energetic resolution has been extracted by means of the already used Breit-Wigner function described in the text. In particular, we have considered that the TH data are nicely described in terms of Eq.(10), once a smearing procedure has been properly applied. The use of more refined approaches, such as a $R$-matrix function, it is not necessary in this context owing to the experimental uncertainties. Assuming a Breit-Wigner shape for the resonance, Eq.(10) has been in fact folded with a Gaussian simulating the response function of the detectors to get the finite-resolution data. Then, in a recursive approach the folded function has been compared with the THM data and the parameter of the original BW modified until the THM data are well reproduced (minimum reduced $\chi^2$).\\
Thus, the TH S(E)-factor at infinite resolution has been evaluated starting from the original analytical expression in Eq.(10), without considering the contribution of the subthreshold level. The BW function describing the $\sim$10 keV resonance as well as the no-resonant contribution of Eq.(10) have been then corrected for the phase-space population effect, penetrability through the Coulomb-barrier, and for the Gamow factor thus allowing us to get the TH S(E)-factor at infinite resolution. The infinite-resolution TH S(E)-factor is shown in Fig.14 as a blue line, while the experimental data at the energy resolution of 31 keV are shown as black-points together with the corresponding smeared function.\\
Fig.15 shows the comparison between the direct data of \cite{Angulo1999} and the THM S(E)-factor at infinite resolution (full blue line) together with its allowed upper and lower values (dashed blue lines). 
The 10 keV THM S(E)-factor is $S(10 keV)_{TH}$ = 3127$\pm$583 (MeV b), the error including statistical, subthreshold subtraction, channel radius and normalization uncertainties. The THM value is in agreement with the extrapolated one reported in \cite{Angulo1993}, 2870$\pm$500 (MeV b). Table V lists the S(E)-factor values in the literature and the ones obtained in this work, while in Fig.16 we compare our THM S(E)-factor with the R-matrix calculation previously described.  Fig.16 shows a very good agreement between two independent approaches, namely, the experimental THM (stars) and the R-matrix calculation performed taking the resonance parameters in the literature (dashed and solid lines for bare-nucleus and screened astrophysical factors, respectively). This fact makes it clear that possible systematic errors might affect direct data in the region where the two data sets from \cite{Youn1991} and \cite{Angulo1993} overlap.
It is important noting the THM S-factor and the R-matrix have the same energy trend; even if the THM relies on direct data for normalization, possible systematic errors would not change our conclusions.\\
\begin{table}[htdp]
\caption{ The $^{10}$B(p,$\alpha$)$^7$Be S(E)-factor values as given in the literature and as obtained in the present work.}
\begin{center} 
\begin{tabular}{llllccc}
\hline 
\hline \\
 {}&${}S(0)$       {}&S(10 keV)                      {}&Approach                                    {} &Ref.                         {} &Year       \\ [1ex] 
 {}&   (MeV b)     {}& (MeV b)                            {}&                                                                 {}&                       {}&              \\ [1.0ex] 
\hline\\
 {}& ------    {}&2200$\pm$600            {}&Direct exp.             {}&\cite{Youn1991}        {}&1991          \\ [1.5ex]   
{}& ------      {}&2870$\pm$500           {}&Direct exp.                                                   {}&\cite{Angulo1993}             {}&1993 \\[1.ex]
 {}&900                        {}&3480                            {}&DWBA                                             {}&\cite{Rauscher1996}      {}&1996    \\[1.ex]                                                                    
 \hline \\
 {}&1116     {}&3105            {}&$R$-matrix                                         {}&present work                       \\ [1.5ex]   
 {}&1247$\pm$312      {}&3127$\pm$583               {}&THM                                     {}&present work                       \\ [1.0ex]                                                                        
 \hline
\hline 
\end{tabular}
\end{center}
\end{table}
 \begin{figure}
\begin{center}
\includegraphics[scale=0.44]{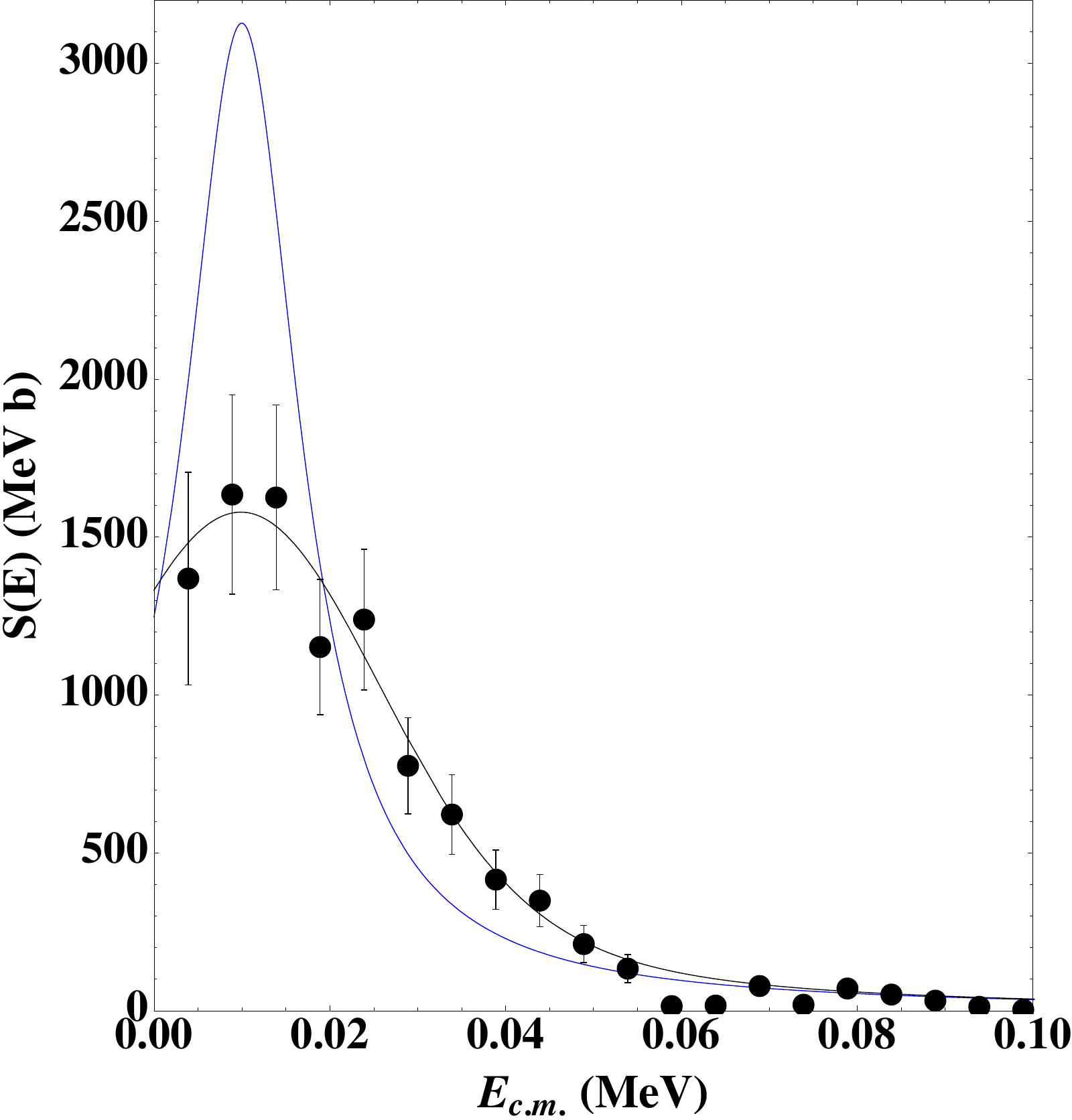}
\caption{(Color online) The experimental TH S(E)-factor (black points with the corresponding uncertainties of Table V) together with its fit. The blue line represents the same TH S(E)-factor after removing the energy resolution effects, as discussed in the text.} 
\end{center}
\end{figure}
 \begin{figure}
\begin{center}
\includegraphics[scale=0.62]{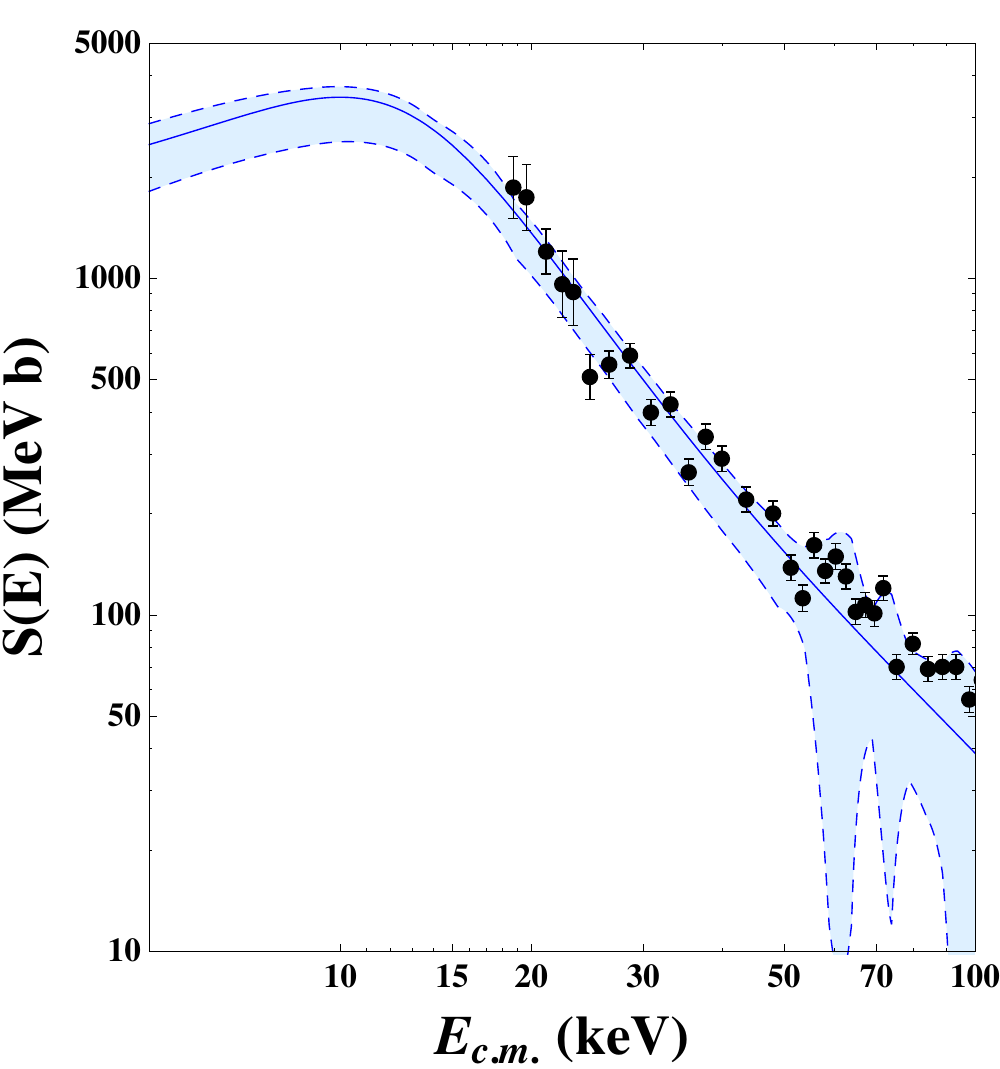}
\caption{(Color online) The TH $^{10}$B(p,$\alpha_0$)$^7$Be S(E)-factor at infinite resolution, together with its allowed upper and lower limits as given by the corresponding uncertainties, is compared with the low-energy direct data of \cite{Angulo1993}. While at energies lower than 30 keV direct data are strongly influenced by electron screening effects, the TH S(E)-factor describes the typical bare-nucleus behaviour.}
\end{center}
\end{figure}
 \begin{figure}
\begin{center}
\includegraphics[scale=0.52]{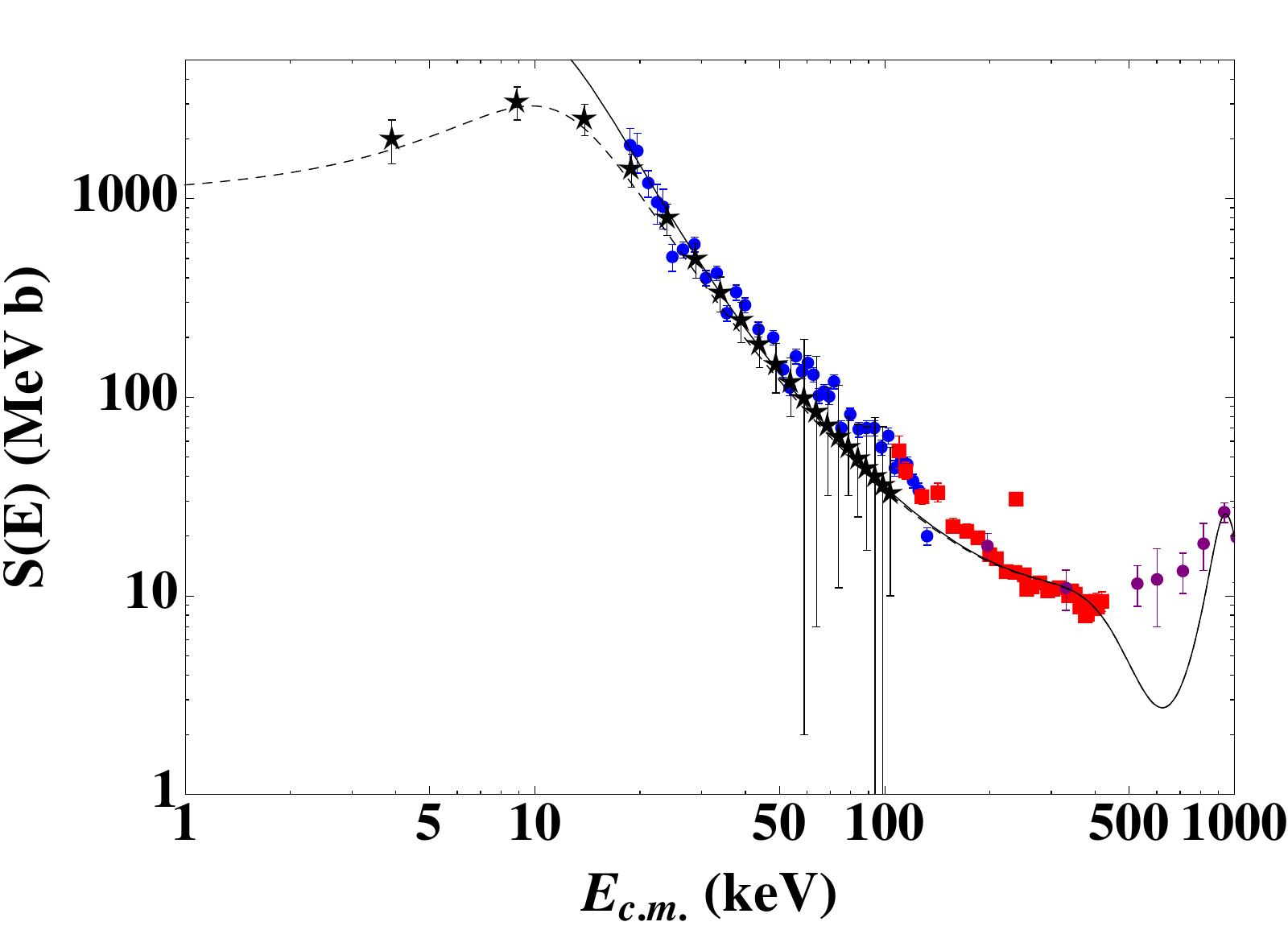}
\caption{(Color online) The THM S(E)-factor (black stars), as given in Table IV, compared with the R-matrix calculation discussed in Section V.B (dashed line) and with the one including electron screening (full line). Red symbols mark the direct data from \cite{Youn1991}, corrected for the factor 1.83 as recommended in \cite{Angulo1993}, blue symbols the data by \cite{Angulo1993} and purple symbols the thick-target data in \cite{atdata79}.}
\end{center}
\end{figure}
 \begin{figure}
\begin{center}
\includegraphics[scale=0.44]{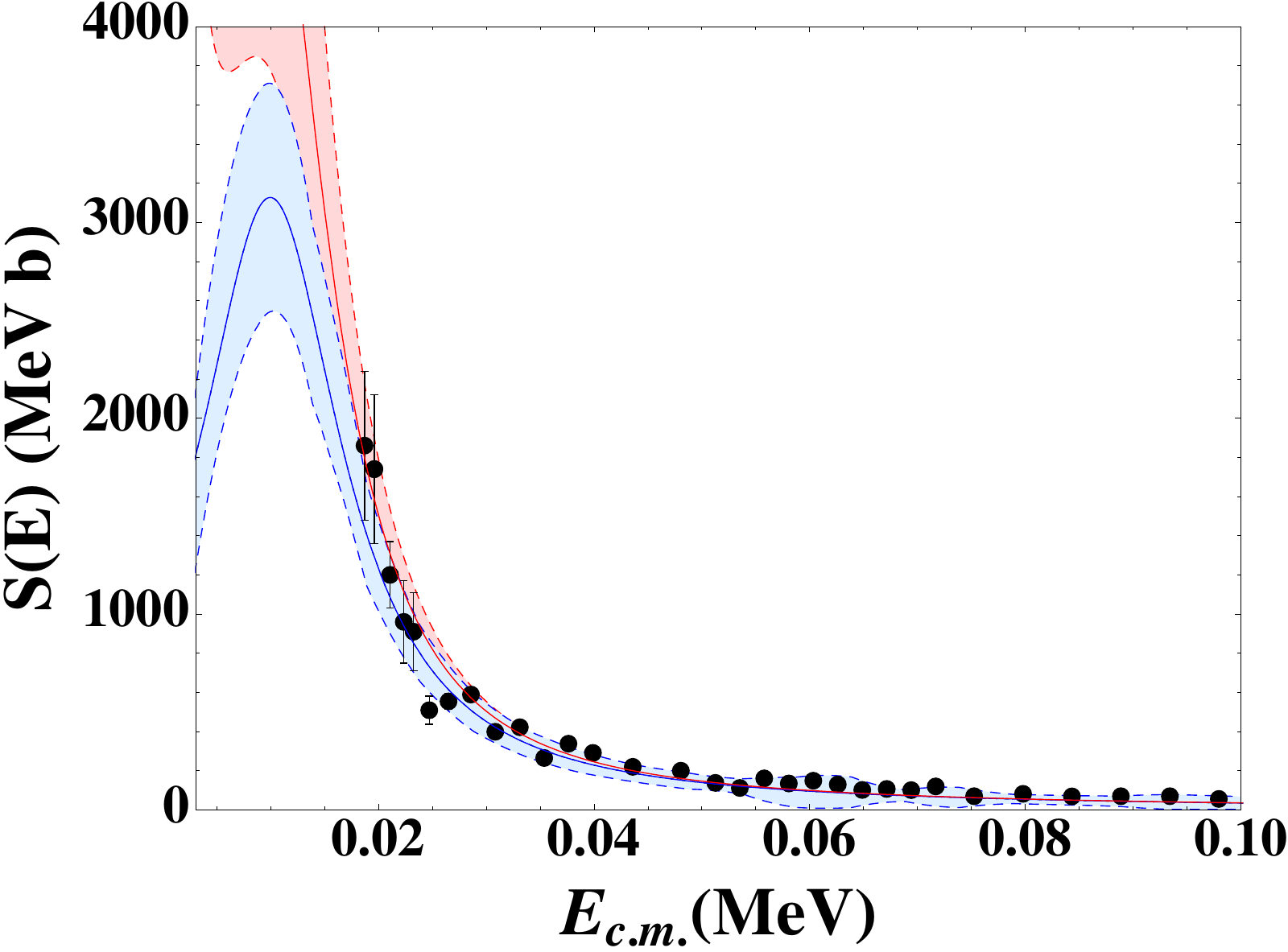}
\caption{(Color online) The bare nucleus THM S(E)-factor at infinite resolution (blue-line) together its upper and lower values (dashed blue line). The low energy data of \cite{Angulo1993} have been then fitted leaving the screening potential U$_e$ as the only free parameter, leading to U$_e$=240$\pm$200 eV. The result is shown as full red line, together with its upper and lower values.}
\end{center}
\end{figure}
In the case of the direct measurements, it must be stressed here that the low-energy cross section evaluations are difficult to be performed, making it necessary to perform extrapolations.\\
It is worth noting that electron screening significantly alters the low-energy trend of the S(E) factor, thus its effect has to be removed before extrapolation to prevent systematic errors. In the $^{10}$B+p case, the adopted enhancement factor assumes the electron screening potential value U$_e$ =  430$\pm$60 eV as deduced from the direct $^{11}$B(p,$\alpha$)$^8$Be S(E)-factor measurement, under the hypothesis of no isotopic dependence of U$_e$ \cite{Assenbaum1987}. \\
Indeed, in the case of the $^{10}$B($p$,$\alpha$)$^7$Be reaction, extrapolation  from high-energy data has been performed, assuming a single level Breit-Wigner function describing the resonance at 10 keV, with parameters from  Ref.\cite{Wiescher1983}.\\
Since the THM provides an independent measurement of the bare nucleus S(E)-factor, the electron screening potential can be extracted by fitting the available low-energy direct data of \cite{Angulo1993} by using the TH bare-nucleus S(E)-factor and the standard expression for the enhancement factor \cite{Rolfs1988,Assenbaum1987,Strieder2001}
\begin{equation}
S_s(E) = [S_b(E)]^{THM} \cdot exp\left(\pi\eta\frac{U_e}{E}\right)
\end{equation}
where $U_e$ is left as the only free parameter in the best fit procedure and $f _{\rm lab}=exp\left(\pi\eta\frac{U_e}{E}\right)$  is the enhancement factor usually introduced to parameterize the rise of the S(E)-factor due to the electron screening effects \cite{Rolfs1988}.\\
As already mentioned, the $[S_b(E)]$$^{TH}$ should show the same trend as the direct   S$_b$(E), except in the ultra-low energy range  where the two data sets should differ due to the effects of electron screening (Fig.15). For such a reason the low-energy direct data of \cite{Angulo1993} have been fitted by using Eq.19, by leaving the electron screening potential $U_e$  as the only free parameter. The procedure returns the result shown in Fig.17 and the value of [$U_e$]$^{TH}$ = 240$\pm$200 eV, where the large error takes into account the uncertainties on the bare-nucleus THM S(E)-factor measured here. The central value is in agreement, within the experimental uncertainties, with the adiabatic limit of 340 eV. Table VI is a summary of the adopted electron screening potential values as given in the literature.
\begin{table}[htdp]
\caption{ Electron screening potential for the boron+proton system. It is worth notice that the $^{10}$B-p direct measurement adopt the same U$_e$ potential deduced from the $^{11}$B-p measurement, while the THM measurement discussed in the text provides an independent U$_e$ determination once the bare-nucleus S(E)-factor has been evaluated.}   
\begin{tabular}{lllcclll}
\hline 
\hline 
  {}&Reaction                                                           {}&$U_e$            {}&Approach                {}&reference    {}&Year\\ [0.0ex]           
  {}&                                                              {}&(eV)                {}&                               {}&                      \\ [1. ex]
\hline
{}&$^{11}$B($p,\alpha_0$)$^8$Be                                   {}&430$\pm$80   {}&Direct exp.      {}&\cite{Angulo1993}   {}&1993 \\[1.5ex]
 {}&              {}&472$\pm$120  {}&THM      {}&\cite{Lamia2012}   {}&2012  \\[1ex]
 {}&$^{10}$B($p,\alpha_0$)$^7$Be                               {}&430$\pm$80        {}&Direct exp.                             {}&\cite{Angulo1993}      {}&1993    \\[1.5ex]

{} &               {}&240$\pm$200     {}&THM    {}&present work            \\[1ex]
\hline
\hline
\end{tabular}
\end{table}

\section{CONCLUSIONS}
 The $^{10}$B(p,$\alpha_0$)$^7$Be reaction has been measured for the first time at the Gamow peak by means of the THM applied to the $^2$H($^{10}$B,$\alpha_0$ $^7$Be)n  QF reaction. The QF reaction mechanism has been quantitatively evaluated by analyzing the relative energy spectra and extracting the experimental momentum distribution for the p-n intercluster motion inside deuteron. Both PWIA and DWBA give the same shape  for the theoretical momentum distribution if one considers neutron momentum values fulfilling the momentum prescription of Eq.(7) \cite{Shapiro1967a}. The experimental THM yield is characterized by the population of three different resonant levels of the intermediate $^{11}$C nucleus, being the 8699 keV one of primary importance for the $^{10}$B($p,\alpha_0$)$^7$Be S(E)-factor determination. In fact,
the Gamow peak for typical boron quiescent burning is centered at 10 keV and coincides with the 8.699 MeV $^{11}$C state , determining a l=0 resonance at such energy. To this aim, energy resolution effects and selection of the events of interest for the THM investigation have been carefully evaluated together with the corresponding uncertainties. In this way the S(E)-factor has been obtained at low energies with no need of extrapolation. By using the high-energy direct data for normalization, the absolute value of the astrophysical factor has been determined, giving, for the first time, a measurement at the corresponding Gamow peak.
In addition, since the THM S(E)-factor does not suffer from electron screening effects, it has been used to evaluate the electron screening potential value needed for the description of the low energy direct data. This represents the first independent measurement of U$_e$ for the $^{10}$B($p,\alpha_0$)$^7$Be reaction, since the adopted one derives from applying the so-called isotopic independence hypothesis for electron screening phenomena. The quoted uncertainties on both S(E) and U$_e$ take into account statistical and systematic uncertainties, for which a careful evaluation has been deeply discussed in the text. The present THM investigation of the $^{10}$B($p,\alpha_0$)$^7$Be reaction leads to $S(10 keV)_{TH}$  = 3127$\pm$583 (MeV b) for the S(E)-factor value in correspondence of the 10 keV resonance, in which the quoted error accounts for statistical, subthreshold subtraction, normalization and channel radius uncertainties. By using the measured bare-nucleus TH S(E)-factor, a value of 240$\pm$200 eV has been deduced for the electron screening potential value, where the large error takes into account the uncertainties on the TH S(E)-factor.\\

\begin{acknowledgments}
This work has been partially supported by the Italian Ministry of University MIUR under the grant ``LNS-Astrofisica Nucleare (fondi premiali)" and RFBR082838 (FIRB2008). It has been partially supported also by the grant LC 07050 of the Czech M\^SMT, grant M10480902 of the Czech Academy of Science, grant LH1101 of the AMVIS project. A. M. M. acknowledges the support by U.S. Department of Energy under Grants No. DE-FG52-09NA29467, No. DE-FG02-93ER40773, and No. DE-SC0004958 and by NSF under Grant No. PHY-0852653.
\end{acknowledgments}

\bibliography{b10_thm}
%
\end{document}